# A Model of an Oscillatory Neural Network with Multilevel Neurons for Pattern Recognition and Computing

**Andrei Velichko *, Maksim Belyaev and Petr Boriskov**

Institute of Physics and Technology, Petrozavodsk State University, 31 Lenina str., 185910 Petrozavodsk, Russia; biomax89@yandex.ru (M.B.); boriskov@petrsu.ru (P.B.)
* Correspondence: velichko@petrsu.ru; Tel.: +7-8142-63-5773

**Abstract:** The current study uses a novel method of multilevel neurons and high order synchronization effects described by a family of special metrics, for pattern recognition in an oscillatory neural network (ONN). The output oscillator (neuron) of the network has multilevel variations in its synchronization value with the reference oscillator, and allows classification of an input pattern into a set of classes. The ONN model is implemented on thermally-coupled vanadium dioxide oscillators. The ONN is trained by the simulated annealing algorithm for selection of the network parameters. The results demonstrate that ONN is capable of classifying 512 visual patterns (as a cell array 3 × 3, distributed by symmetry into 102 classes) into a set of classes with a maximum number of elements up to fourteen. The classification capability of the network depends on the interior noise level and synchronization effectiveness parameter. The model allows for designing multilevel output cascades of neural networks with high net data throughput. The presented method can be applied in ONNs with various coupling mechanisms and oscillator topology.

## 1. Introduction

Hypotheses about the functional importance of synchronization for information processed by the brain were put forward long ago [1,2] and its experimental discovery [3,4] encouraged the creation of neural networks models with oscillatory dynamics [5–7] and neuromorphic algorithms of image processing based on synchronization [8–11].

Research on ONN is mainly based on phase oscillator models. Such models, primarily the Kuramoto model [12], have been very useful for studying systems of various topology with a large number ($N > 10^3$) of oscillators, where not only global synchronization but the synchronization by middle field [13], mode of quasi-synchronization [14] and even chimeras synchronization [15] are feasible.

Recently, the problem of pattern recognition in ONN has been intensively studied [16,17], and two main global synchronization methods have been outlined: frequency-shift [10] and phase-shift [11] keying methods. Frequency encoding, based on synchronized frequency shift [10], allows usage of an oscillator star configuration and $N$ couplings; however, the disadvantage of this method is that only it stores a single pattern. The phase-shift keying method of encoding [11] enables storage of more than one pattern by certain combinations of phase shift at the same weight ONN matrix. However, this method has the following drawbacks: $N^2$ couplings with tunable weights and a two-stage procedure of pattern recognition.

Another class of ONN is based on relaxation oscillators that generate multiple pulses of short duration and fixed amplitude. Such pulses (spikes) can code information at pulse-repetition frequency. An important distinguishing feature of a pulse type ONN from classic spiking neural networks (SNN) is a self-oscillating mode of ONN neuro-oscillators. This mode is not indicative for SNNs and real (biological) neurons, which are characterized by a forced response through generation of a single spike or their group, when the neuron threshold level is achieved. However, ONNs are fascinating due to the simplicity of the hardware implementation as available advanced micro- and nano-electronic self-oscillators ensure the compact size and energy efficiency of the circuit.

Pulse type ONNs with a multi-frequent spectrum of periodic oscillation feature a specific mode of multiple frequency synchronization, or in other words, a high order synchronization effect [18] (harmonic locking [6]). We demonstrated this effect experimentally by using thermally-coupled vanadium dioxide ($VO_2$) oscillators [19]. In relaxation oscillators with $VO_2$-based film elements, electric self-oscillations are activated by the effect of electric switching governed by metal—insulator transition (MIT) [20–23]. The processing speed of $VO_2$ devices switching which amounts to ~10 ns [24] and manufacturing technology that allows switching elements to be created with high levels of nano-scalability make $VO_2$-switch based oscillators the perfect objects for research on neuro-oscillators to solve cognitive technology problems [25–28]. Relaxation oscillators with high order synchronization effects can be realized by using electric coupling [25–28] and by using not only $VO_2$-switches, but any other switching elements such as thyristors [29], tunneling diodes [30], resistive memory cells [31], and spin torque oscillator [16,32].

In this paper, we studied the ONN of thermally-coupled $VO_2$-oscillators and present a general concept of visual pattern recognition based on high order synchronization effects. We used the multiplicity of synchronous states to extract object classes by using a single output oscillator (compared to, for example, an array of oscillators at the output [10,11]). The concept of a multi-level neuron allows for using a smaller number of output neurons (oscillators) to implement the more complex cognitive functions of the neural network. We developed a set of special metrics [19,33], such as the high-order synchronization value and the synchronization effectiveness value. Compared to the neural network presented in [33], this network is able not only to memorize and classify patterns, but also to perform logical operations, computer calculations and emulate other functions of artificial intelligence systems.

## 2. Materials and Methods

### *2.1. Oscillator Circuit and Method of ONN Organization*

The basic element in the studied ONN is an oscillator implemented on the circuit of a relaxation generator (Figure 1) based on a $VO_2$-switch. We described the process of fabrication and electric I–V characteristics of an electric $VO_2$-switch in detail in [19]. I-V characteristics are well approximated by a piecewise linear function (see Appendix A.1) that has two conduction states (low-resistance and high-resistance) and a region of negative differential resistance (NDR). These switches may be used in the circuit of a relaxation generator with power supply $I_p$ holding the operation point in the NDR region of the I-V characteristic and with capacity $C$, connected to the switch in parallel. In addition, a source of noise $U_n$ models the circuit's interior or exterior noises, such as current noise of a switch [34]. The oscillator's output signal is current $I_{sw}(t)$ flowing through the $VO_2$-switch, which is used to calculate the synchronization level of two oscillators (see Section 2.3). The current signal directly determines the effect of thermal coupling inside the network.

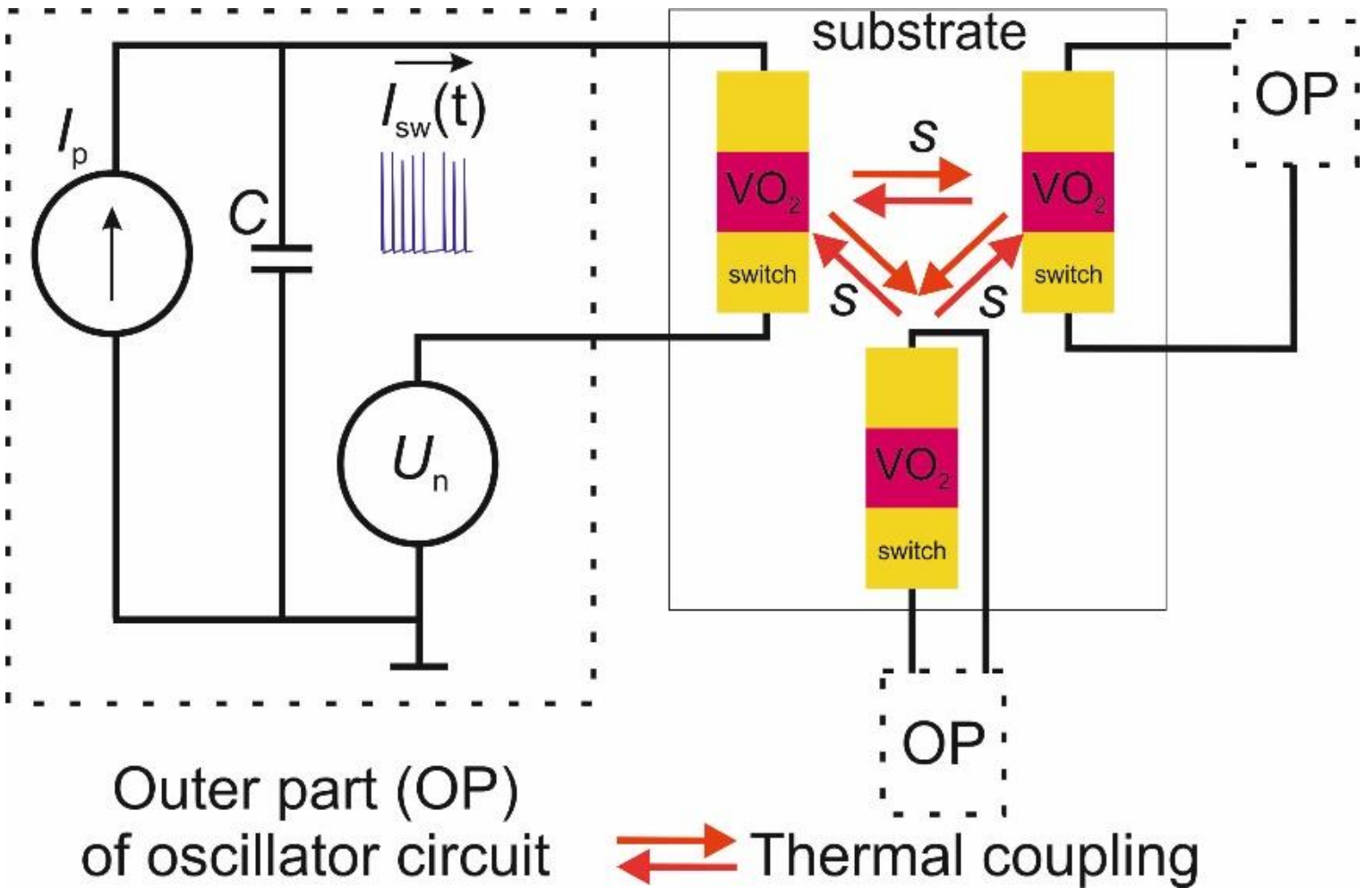

**Figure 1.** The oscillator circuit and example of oscillators' interaction via thermal coupling of $VO_2$ switches. $I_{sw}(t)$—current signal in a $VO_2$ switch, that leads to its Joule heating resulting in thermal impulses that spread along the substrate; $U_n$—source of noise, $s$—thermal coupling strength.

We applied thermal coupling to connect oscillators in a network. The presence of the thermal coupling mechanism between two $VO_2$ oscillators was convincingly demonstrated by us in the experiment [19], and it is based on the mutual thermal effect of switches due to their Joule heating and the dependence of the threshold voltage $U_{th}$ on temperature. In the pre-threshold mode, when the $VO_2$ temperature of the switch channel is close to the MIT temperature $T_{th}$ ~ 340 K [23], the external thermal influence can "push" it to switch, which is equivalent to lowering the threshold voltage by the value of $s$, the thermal interaction force. The switching causes an even higher temperature rise in the switch, which leads to the appearance of a temperature pulse in the surrounding space, which propagates as a temperature wave. In the oscillator circuit, the thermal effect of the switches on each other occur in the mode of repetitive pulses initiated by self-oscillations of oscillator currents, which ultimately leads to their synchronization [19]. The value of $s$ can be controlled experimentally, for example, by varying the distances between the switches or by varying the parameters of the external circuit [19].

The choice of this coupling type is determined by the simplicity of a computing model when oscillators are electrically separated from each other, unlike in capacitive or resistive couplings [21,22,25,28]. An example of an oscillators' interaction via thermal coupling of $VO_2$ switches is shown in Figure 1. A detailed presentation of the mathematical model of thermally coupled relaxation oscillators is given in Appendix A.1.

The heat assisted model we used completely describes the experimental data previously observed [19,35,36]. Indeed, the speed of propagation of a thermal wave can be taken into account, but this is not in the scope of the current paper because we operated at low oscillation frequencies where the time delay is insignificant. We considered the radius of thermal interaction $R_{TC}$ in [19], therefore, in this model we limited ourselves to interaction with the nearest oscillators. Consideration of more complex cases can be done in future publications.

We used the concept of one-way thermal coupling of oscillators in the numerical model. This can be physically implemented when a resistance heater is used as a connecting element in one of the circuits instead of a switch [19].

For numerical modeling we used the following parameters as I-V characteristics ($U_{th}$ = 5 V, $U_h$ = 1.5 V, $U_{bv}$ = 0.82 V, $R_{off}$ = 9.1 kΩ, $R_{on}$ = 615 Ω). In this circuit, capacity is a constant parameter $C$ = 100 nF, and its value significantly determines the frequency range of oscillator operation [37] and its natural frequency $F_0$. In our case, the frequency range was 165 Hz ≤ $F_0$ ≤ 1266 Hz at the range of feeding currents 550 μA ≤ $I_P$ ≤ 1061 μA.

## *2.2. ONN Structure*

The studied ONN consists of an input pattern, which is transferred as a 3 × 3 matrix on the layer of processing neurons consisting of 9 oscillators, and of an output layer with only one oscillator (output neuron) No.10 (see Figure 2).

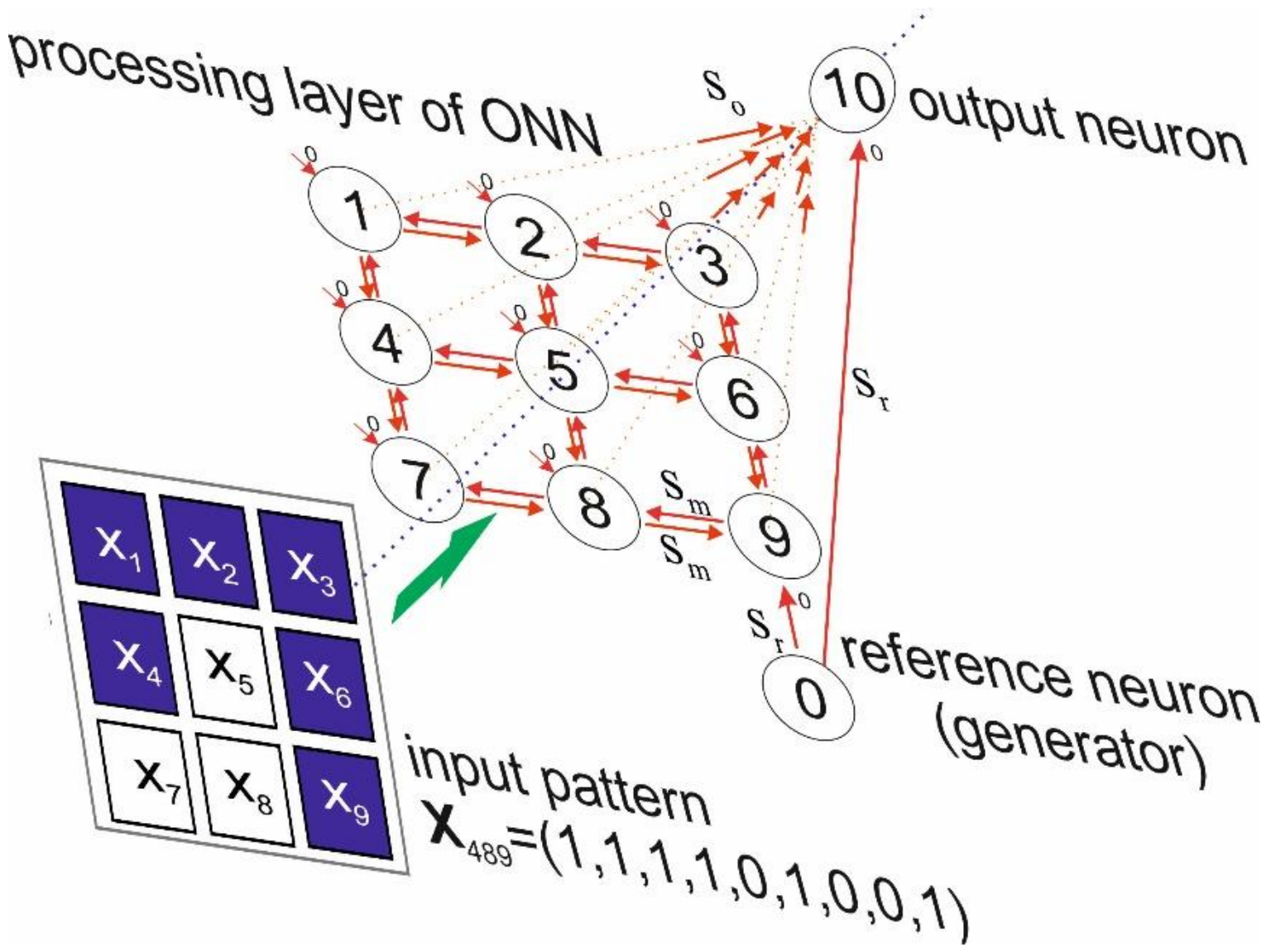

**Figure 2.** ONN organization circuit for pattern recognition as a matrix of 3 × 3 elements. Digits indicate the sequence numbers of oscillators.

The magnitude of the effect of the *i*-th oscillator on the *j*-th oscillator is determined by coupling strength $s_{i,j}$, that is set by the following matrix:

$$S=\begin{bmatrix} s_{0,0} & \cdots & s_{0,10} \\ \cdots & \cdots & \cdots \\ s_{10,0} & \cdots & s_{10,10} \end{bmatrix}=\begin{bmatrix} 0 & s_r & s_r & s_r & s_r & s_r & s_r & s_r & s_r & s_r & s_r \\ 0 & 0 & s_m & 0 & s_m & 0 & 0 & 0 & 0 & 0 & s_o \\ 0 & s_m & 0 & s_m & 0 & s_m & 0 & 0 & 0 & 0 & s_o \\ 0 & 0 & s_m & 0 & 0 & 0 & s_m & 0 & 0 & 0 & s_o \\ 0 & s_m & 0 & 0 & 0 & s_m & 0 & s_m & 0 & 0 & s_o \\ 0 & 0 & s_m & 0 & s_m & 0 & s_m & 0 & s_m & 0 & s_o \\ 0 & 0 & 0 & s_m & 0 & s_m & 0 & 0 & 0 & s_m & s_o \\ 0 & 0 & 0 & 0 & s_m & 0 & 0 & 0 & s_m & 0 & s_o \\ 0 & 0 & 0 & 0 & 0 & s_m & 0 & s_m & 0 & s_m & s_o \\ 0 & 0 & 0 & 0 & 0 & 0 & s_m & 0 & s_m & 0 & s_o \\ 0 & 0 & 0 & 0 & 0 & 0 & 0 & 0 & 0 & 0 & 0 \end{bmatrix} \quad (1)$$

The layer of processing neurons consists of a 3 × 3 oscillator matrix connected by similar couplings ($s_{i,j} = s_{j,i} = s_m$, where *i*, *j* are the numbers of neighboring oscillators). Neighboring oscillators are only connected by horizontal and vertical lines. So, the central oscillator No.5 has four couplings in the matrix, the corner oscillators have two couplings, and the oscillators in the center of the edges have three couplings (with the central oscillator and two corner ones) and they all unidirectionally affect oscillator No.10 (output neuron) in the output layer ($s_{i,10} = s_o$ and $s_{10,i} = 0$, where $i = 1 \ldots 9$). Importantly, there is the reference oscillator No.0 in the circuit and the synchronization order of all other oscillators is measured against this oscillator. Oscillator No.0 (Figure 2) unidirectionally affects all other oscillators ($s_{0,i} = s_r$, and $s_{i,0} = 0$, where $i = 1 \ldots 10$).

The input pattern (see Figure 3) is transferred to the processing layer by selection of the feeding currents of the oscillator matrix:

$$I_{p_i} = \begin{cases} I_{ON}, \text{ if } x_i = 1 \\ I_{OFF}, \text{if } x_i = 0 \end{cases} \quad (2)$$

where $i = 1 \ldots 9$, $x_i$ are the coordinates of the input vector X = ($x_1$, …, $x_9$), that correspond to white ($x_i$ = 0) and blue ($x_i$ = 1) colors of pattern squares.

Transfer of the pattern to the intermediate layer causes the change in the oscillators' feeding currents, and in turn, it leads to the change in synchronization state for all oscillators (No.1–10). The synchronization order $SHR_{0,10}$ between the reference oscillator No.0 and the oscillator in the output layer No.10 serves as the control value. The values of the feeding currents for these oscillators are fixed and may differ from currents in the matrix, therefore they are indicated as $I_0$ and $I_{10}$ (see Figure 3).

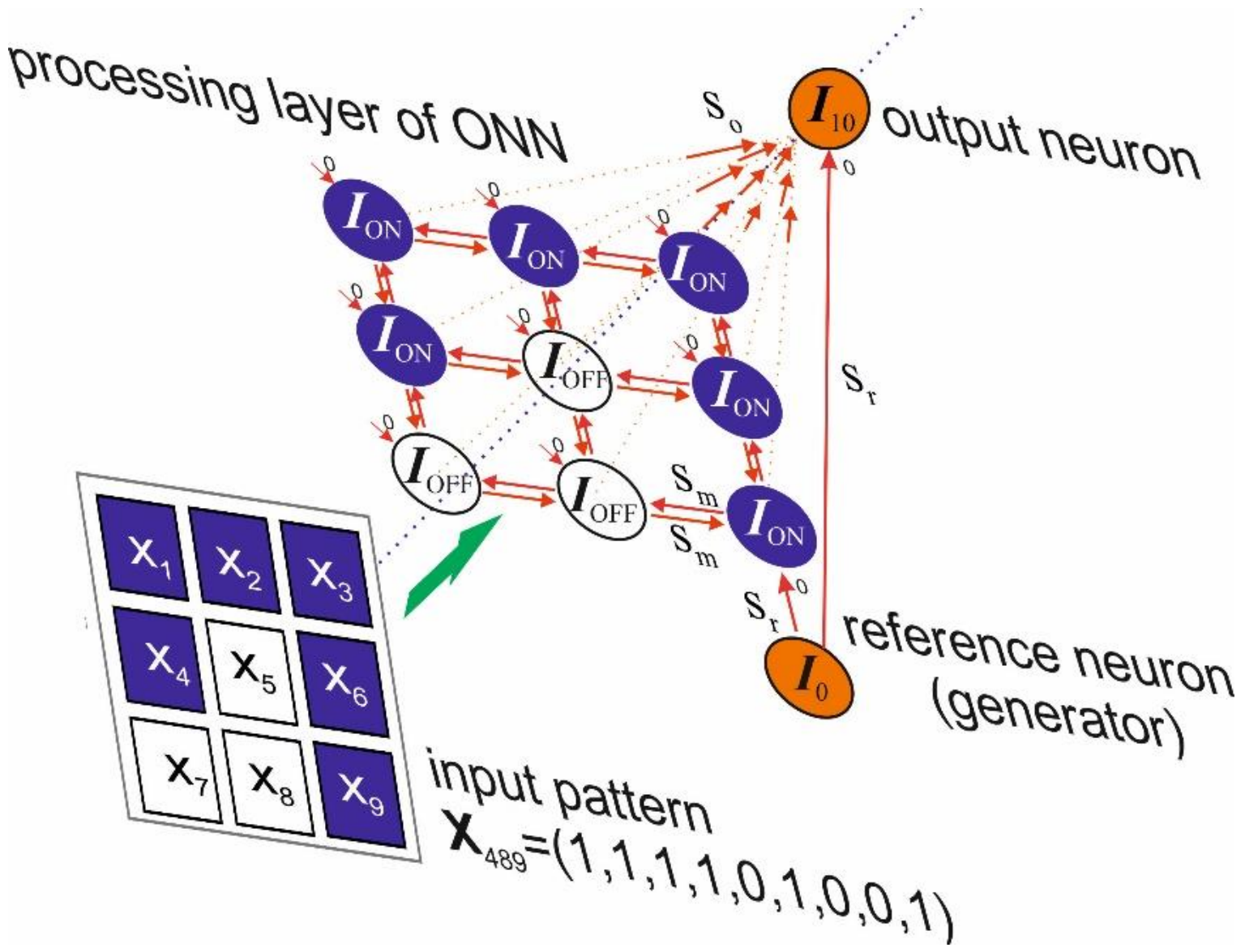

**Figure 3.** Principle of a pattern transfer from the input layer to the oscillator matrix via setting of their currents ($I_{OFF}$ is for white squares, $I_{ON}$ is for blue squares). Separate colors are used for oscillators No.0 and No.10 and for their currents $I_0$ and $I_{10}$, respectively.

## *2.3. Method of Synchronization Order Definition*

In this section, we present the method of calculating a family of metrics [19] to determine the high order synchronization of two oscillators. A family of metrics has two basic parameters: the ratio of subharmonics (SHR) and synchronization effectiveness $\eta$. In a general case, it is possible to use the concept of subharmonics ratio between oscillators *i* and *j*, which is defined [18] as a simple fraction

$$SHR_{i,j} = k_i : k_j \quad (3)$$

where $k_i$ and $k_j$ are harmonics order of oscillators at the common frequency of their synchronization $F_S$.

In other words, Equation (3) uses spectral approaches to describe synchronization of the higher order $k_i$:$k_j$, when the following relation is observed:

$$F_S = k_i \cdot F_i^0 = k_j \cdot F_j^0 \quad (4)$$

where $F_i^0$, $F_j^0$ are frequencies of main harmonics.

As an example, Figure 4 shows the qualitative spectra of two oscillators at $SHR_{i,j}$ = 2:7

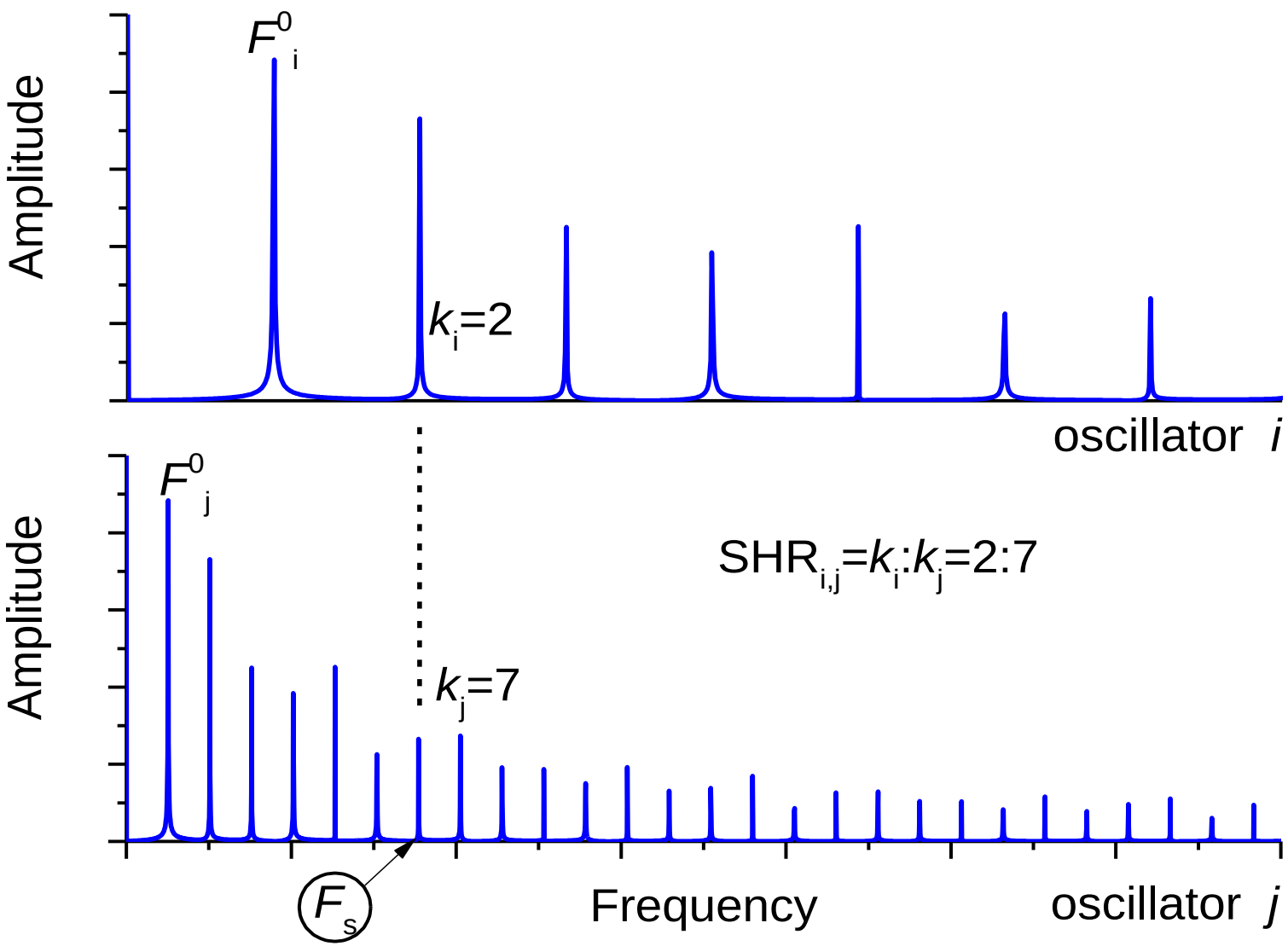

**Figure 4.** Qualitative spectra of two coupled synchronized oscillators with parameter $SHR_{i,j}$ = 2:7.

In addition, Figure 4 shows the total synchronization frequency $F_s$ and subharmonics numbers at this frequency $k_i$ = 2 and $k_j$ = 7. Usage of signal spectra for estimating the magnitude of $SHR_{i,j}$ is not effective because the signals might not have a strictly periodic sequence and when noise is added to the system, the spectral lines broaden. Below, we will describe the mathematical procedure that estimates the value of $SHR_{i,j}$ without the use of spectral characteristics but using a current signal and array LE. Array LE stores information on the position of the current pulse leading edges (see Figure 5).

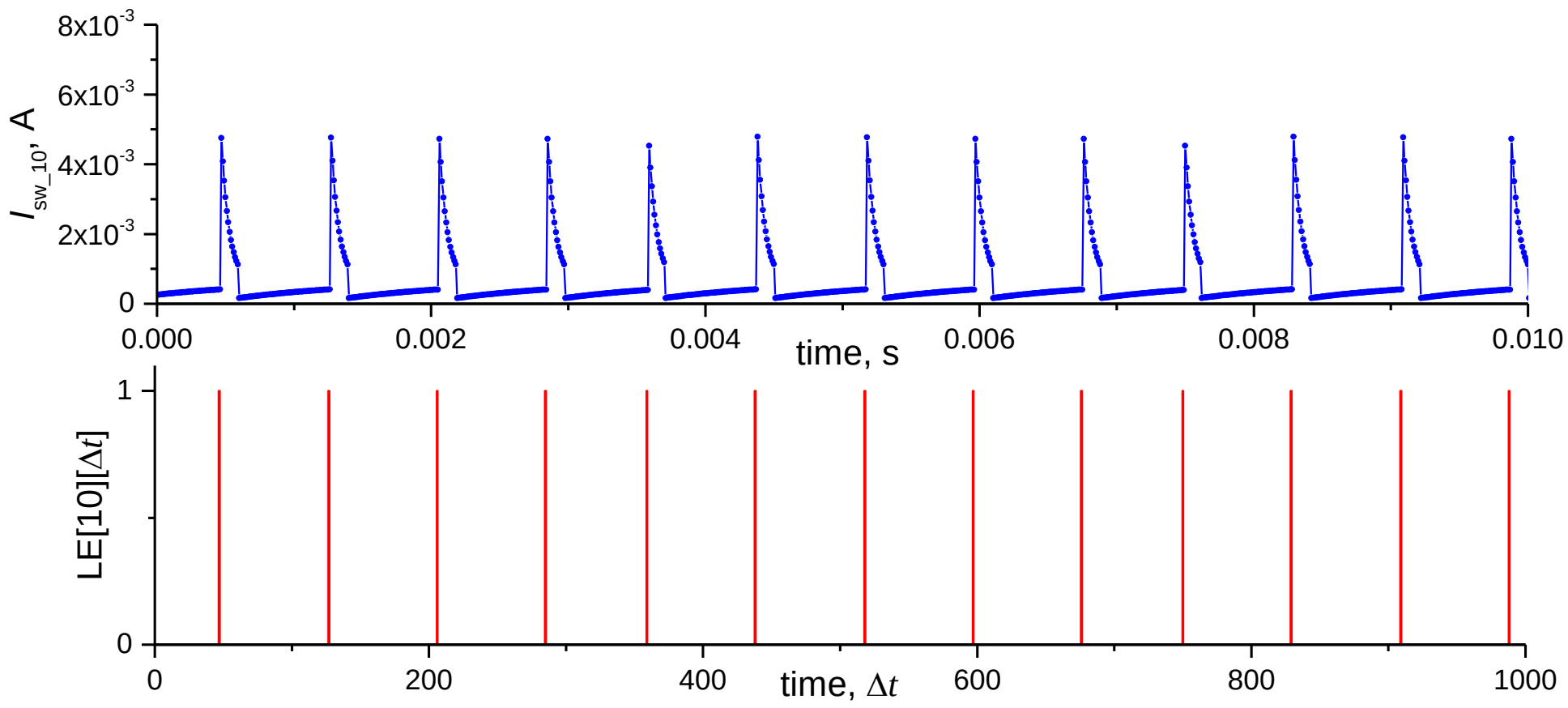

**Figure 5.** Oscillogram of oscillator current No. 10 and the corresponding array of positions of the leading edges of the current pulse LE [10] [Δt]. Δt—calculation time interval (Appendix A.1).

Figure 6 shows arrays LE[*i*] and LE[*j*] for two oscillators that correspond to the same ratio of the basic frequencies as shown in the spectra in Figure 4. The distance between two nearest phase-locked pulses is denoted as $T^z_s$—the period of synchronization (where *z* is a conditional number of periods $T_s$).

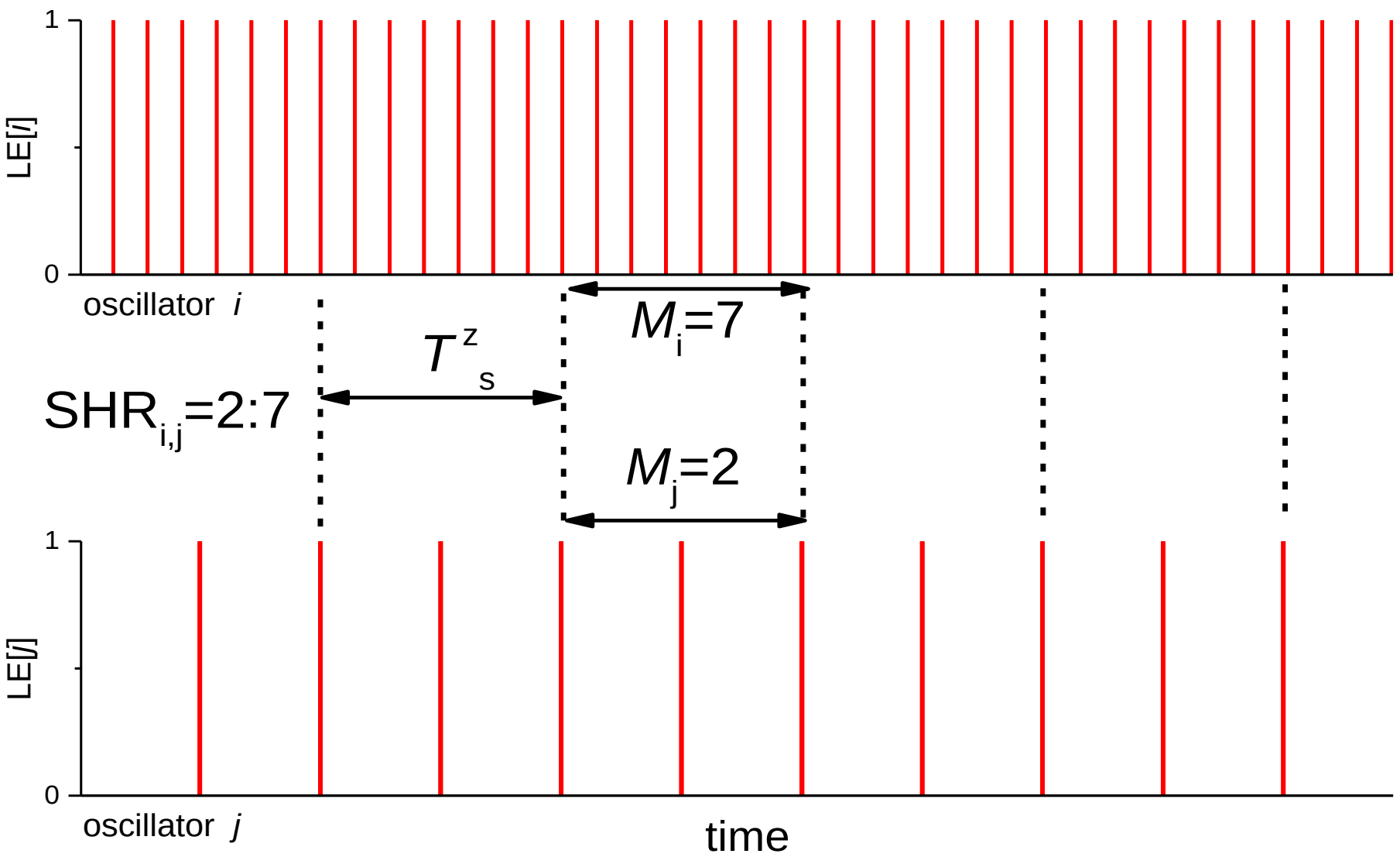

**Figure 6.** Arrays LE[*i*] and LE[*j*] for two oscillators.

Therefore, $SHR_{i,j}$ may be estimated using a phase-locking method:

$$\mathrm{SHR}_{\mathrm{i,j}} = M_{\mathrm{j}} : M_{\mathrm{i}} \quad (5)$$

where $M_i$ and $M_j$ are the number of signal periods falling into the synchronization periods $T^z_s$ of two oscillators. Formula (5) can be easily obtained from Formula (3) taking into account Formula (4) and the following ratio ($T_s = M_i \cdot T_i^0 = M_j \cdot T_j^0$, $F_i^0 = 1/\ T_i^0$, $F_j^0 = 1/T_j^0$).

In general, especially when a system behaves erratically, synchronization periods differ and spread in $T^z_s \neq T^{z+1}_s$ and the values of $M_i$ and $M_j$ may change within one oscillogram (see Figure 7).

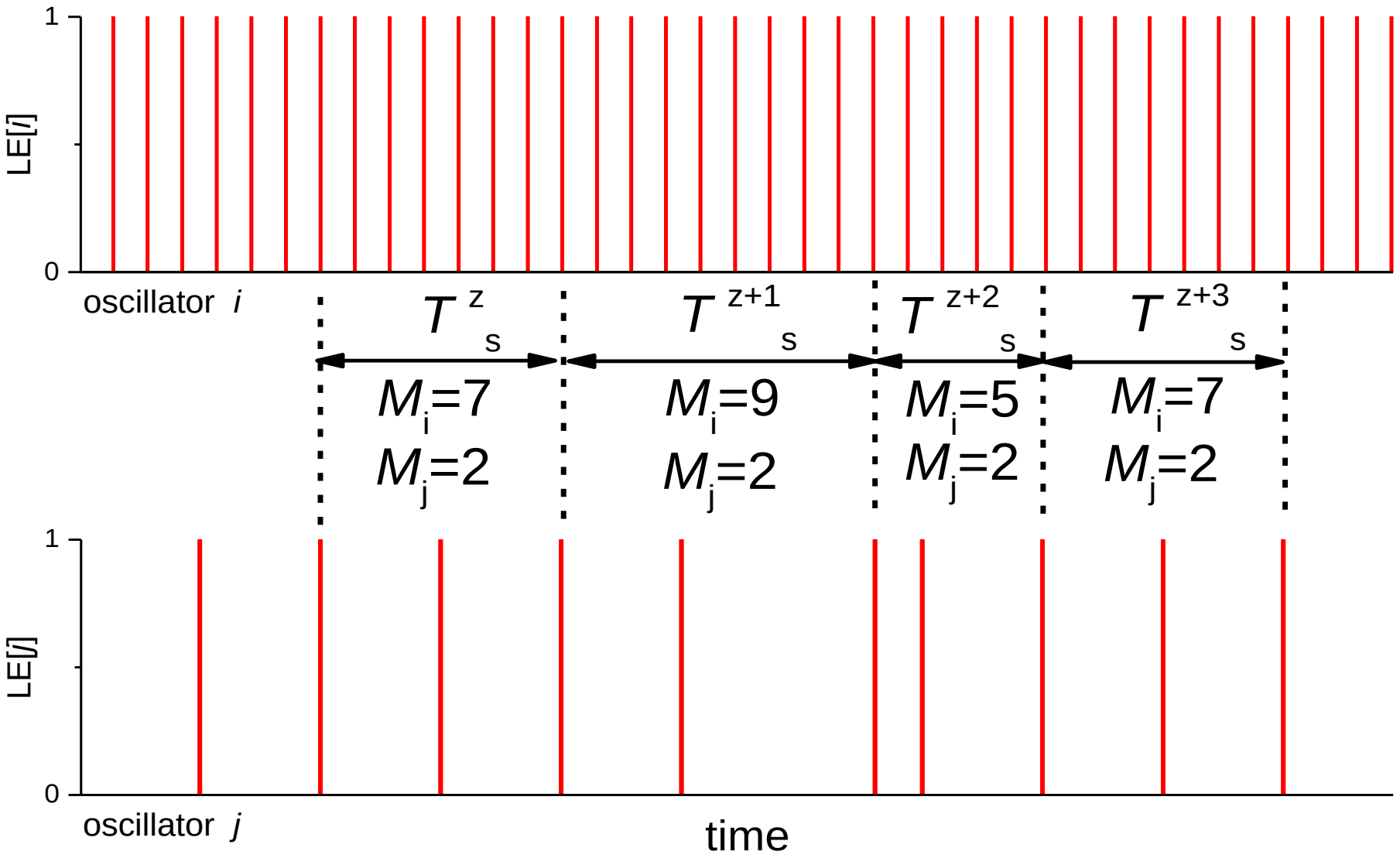

**Figure 7.** Arrays LE[*i*] and LE[*j*] for two oscillators with a non-constant period of synchronization $T^z_s$.

Various values of synchronizations $SHR_{i,j}$ may occur within one oscillogram. To determine which $SHR_{i,j}$ value prevails, it is necessary to find the occurrence probabilities $P(M_j{:}M_i)$ for each pair ($M_i{:}M_j$) that is present in the whole oscillogram and to select the pair with the maximum value of $P = P_{max}(M_j{:}M_i)$. Then, the final value of $SHR_{i,j}$ will be written as:

$$\mathrm{SHR}_{\mathrm{i,j}} = M_{\mathrm{j}} : M_{\mathrm{i}}, \text{if } P = P_{\max}(M_{\mathrm{j}} : M_{\mathrm{i}}) \quad (6)$$

To find the probabilities $P(M_j{:}M_i)$, we can count how many times $NP(M_j{:}M_i)$ the given pair appeared within the whole oscillogram of the oscillator *i*, multiply by the number of periods in it ($M_i$) and divide by the total number of all oscillations periods in the given signal ($N_j$). Thus, for $P(M_j{:}M_i)$ we obtain:

$$P(M_{\mathrm{j}} : M_{\mathrm{i}}) = 100\% \cdot NP(M_{\mathrm{j}} : M_{\mathrm{i}}) \cdot M_{\mathrm{i}} \,/\, N_{\mathrm{i}} \tag{7}$$

where $N_i$ is the total number of periods in the oscillogram of oscillator *i*.

It is convenient to present the probabilities $P(M_j{:}M_i)$ as a histogram where the values are positioned in the descending order of the magnitude *P*. For example, for the oscillogram section in Figure 7 the following histogram can be built. The histogram in Figure 8 is calculated by Formula (7), when the pairs occur the following number of times, *NP*(2:7) = 2, *NP*(2:9) = 1, *NP*(2:5) = 1, and the total number of periods is $N_i = 28$ (in real calculations, $N_i$ was in the range of 1000–3000 for greater accuracy, see Appendix A.2).

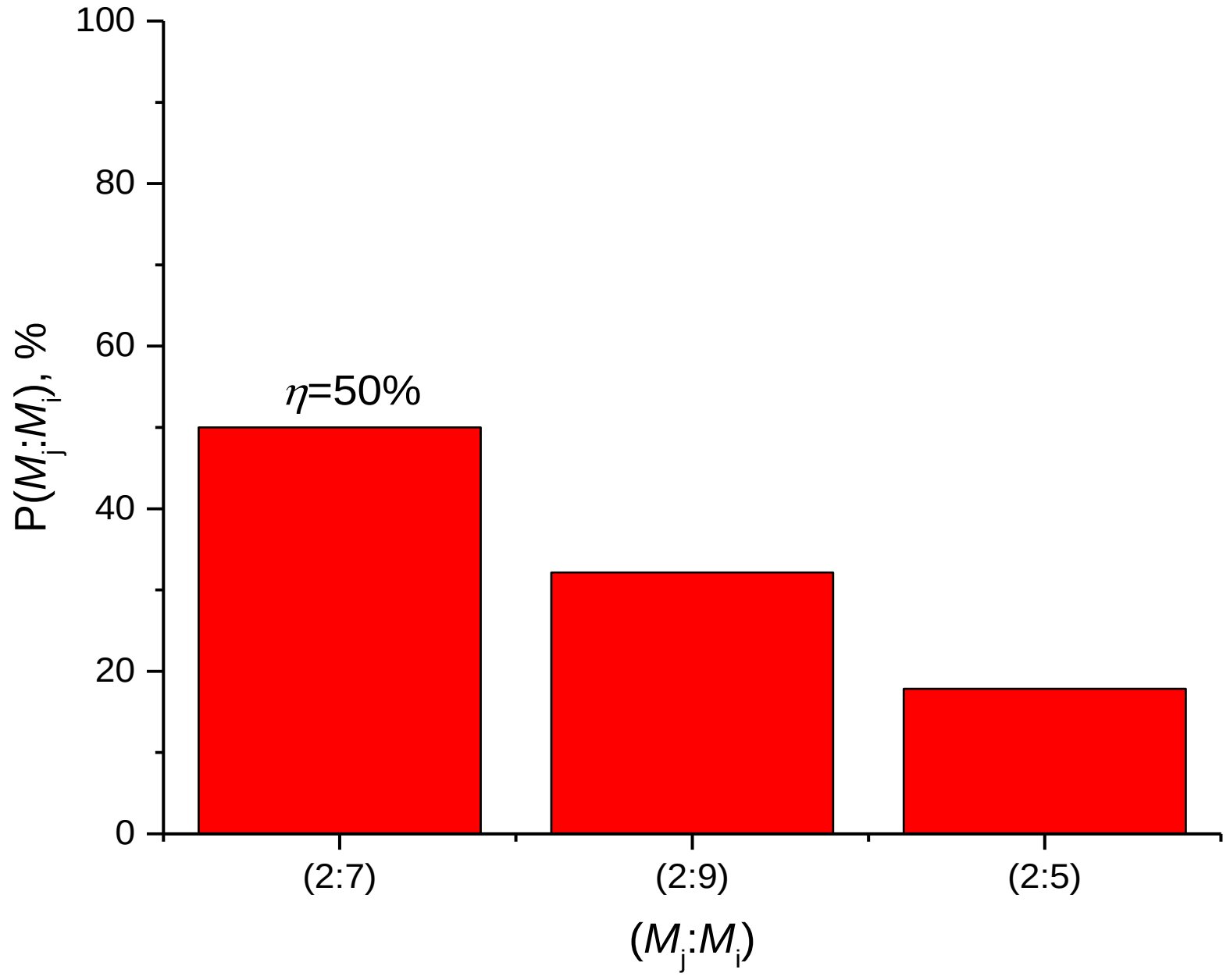

**Figure 8.** Histogram of probabilities distribution $P(M_j, M_i)$ calculated by using Formula (11) for signals LE, shown in Figure 7.

For $SHR_{i,j}$, the parameter of synchronization effectiveness $\eta$ is defined as the maximum probability $P_{max}(M_j{:}M_i)$:

$$\eta = P_{\max}(M_{\mathrm{j}} : M_{\mathrm{i}}) \tag{8}$$

Therefore, we define the synchronization calculated above as $SHR_{i,j}$ = 2:7 with effectiveness of $\eta$ = 50%.

The family of metrics ($SHR_{i,j}$, $\eta$) allows sufficient determination of the synchronization state of two oscillators and calculation of the distance between the states, i.e., the difference between the synchronization degree. This property allows the use of metrics in an oscillator neural network training, pattern recognition systems and artificial intelligence [8–11].

Depending on the task, for example, the network training for data coding and pattern recognition, the problem of the presence or absence of synchronization $SHR_{i,j}$ can be solved by formally setting the synchronization effectiveness threshold $\eta_{th}$, so

$$\text{signals are} \begin{cases} \text{synchronized, if } \eta \geq \eta_{\mathrm{th}} \\ \text{not synchronized, if } \eta < \eta_{\mathrm{th}} \end{cases} \tag{9}$$

In the majority of cases, we set $\eta_{th}$ = 90%, meaning the signals are synchronized if 90% of their durability have a certain synchronization pattern. For the network training, this parameter can be chosen within a selected range, and it is one of the important parameters of the network adjustment [33].

The main technical problem we faced, was the problem of defining the synchronization order between the reference oscillator No.0 and the oscillator of the output layer No.10 characterized by the value $SHR_{0,10}$:

$$\mathrm{SHR}_{0,10} = k_0 : k_{10} \sim k_0 / k_{10} \tag{10}$$

The same value of $SHR_{0,10}$ can be expressed in several ways: as a ratio, a simple fraction or a real number, for example, $SHR_{0,10}$ = 10:3 = 10/3 = 3.33. Later, we will use this property to present the results more vividly.

Parameter $SHR_{0,10}$has the properties of an output neuron while the reference neuron may be considered as a master generator to which we calculate synchronization of other network neurons.

Two parameters SHR and $\eta$ are used as the main metrics for evaluating the degree of two oscillators' synchronization and are applied in the algorithm of ONN training.

The current oscillograms $I_{sw}(t)$ of oscillators No.0–10 were calculated simultaneously and contained ~250,000 points with the time interval $\Delta t$ = 10μs (see Appendix A.1). Then, the oscillograms were automatically processed.

The switch parameters were unchanged in numerical simulation (see Section 2.1), while current intensities $I_{p_i}$ ($I_{ON}$, $I_{OFF}$, $I_0$, $I_{10}$), coupling strength constants $s$ ($s_r$, $s_m$, $s_o$), noise amplitude $U_n$ and $\eta_{th}$ varied.

### *2.4. Pattern Classifier Implementation and Problem Definition*

A "black-and white" pattern was used as an input test pattern presented as a 3 × 3 matrix (without gradation of gray color, 3 by 3 pixels). The form of the pattern can be unequivocally defined by the input vector $X_n = (x_1, \ldots, x_9)$ where each cell may take the value $x_i$ = 0 (white color) or $x_i$ = 1 (blue color), and $n$ is the number of the vector equal to the decimal value of the coordinates presented as a binary sequence. For example, Figures 2 and 3 show an input pattern that corresponds to the vector $X_{489}$ = (1,1,1,1,0,1,0,0,1). The total number of patterns (vectors) $n$ in the input layer matrix is $2^9$ = 512 ($X_0$ … $X_{511}$). Presuming that the pattern processing layer together with the output layer has certain symmetry, a set of 512 vectors $X_n$ may be divided into 102 classes $C_j$, where $j$ is the number of classes ($j$ =1..102) (see Figure 9). The complete list of classes and their elements is described in Supplementary Materials (Data1.txt).

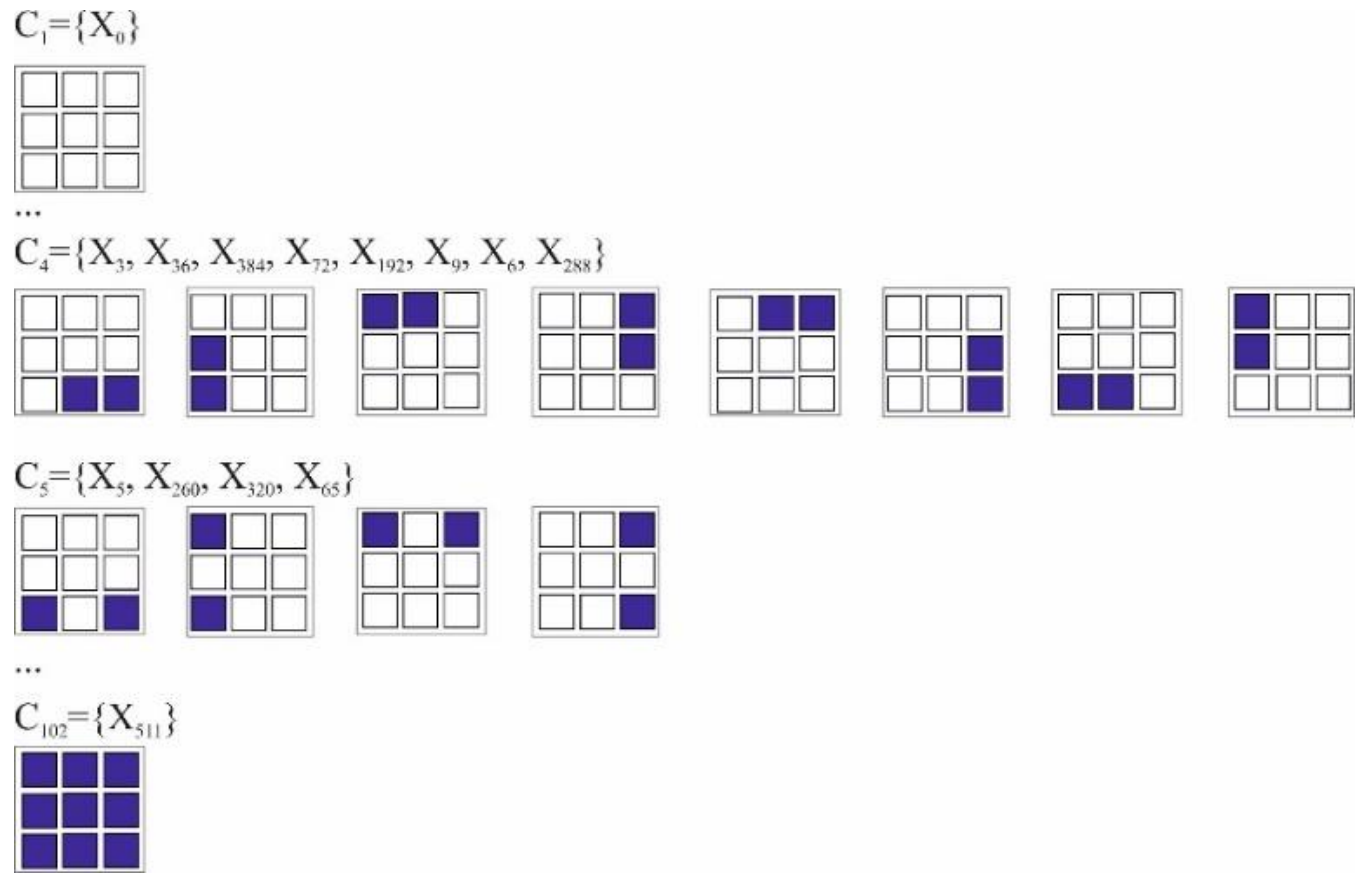

**Figure 9.** Symmetry-based distribution of 512 figures (vectors $X_n$) into 102 classes $C_j$.

The principle of figure distribution into classes is as follows: each class $C_j$ consists of a lot of figures (vectors) that have the same number of blue (white) cells and rotation-reflection axial symmetry of the 4th order (symmetry at rotation by 90°).

Mirror-rotation axial symmetry of the 4th order supposes an association of all figures in a separate class in the mirror operation in relation to the central columns (horizontally and vertically) and also at rotation by 90° (see the example of figures transformation for class $C_4$ in Figure 10).

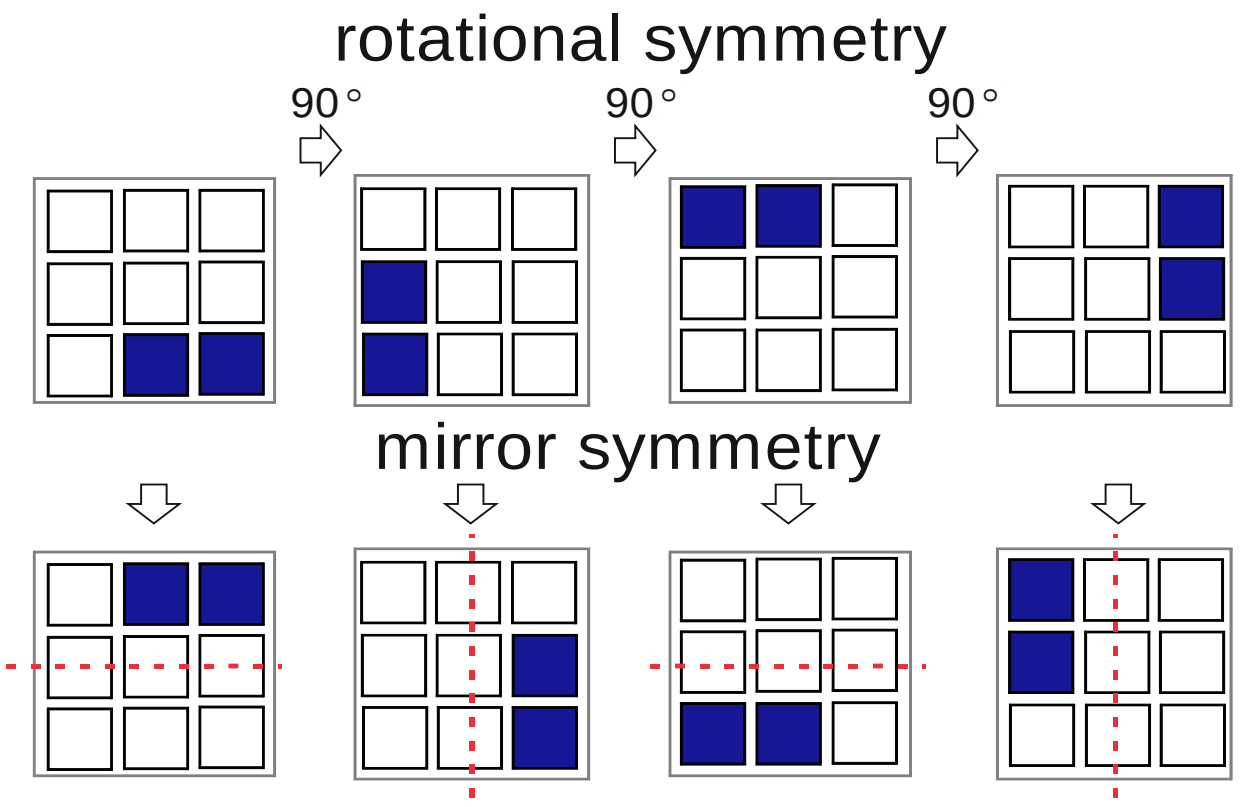

**Figure 10.** Mirror-rotation axial symmetry of the 4th order in class $C_4$.

Figures from one class impose the same effect on the neural network and have the same output value $SHR_{0,10}$. Distribution into classes allows us to find figures that at any current values ($I_{ON}$, $I_{OFF}$, $I_0$, $I_{10}$) and coupling strength constants ($s_r$, $s_o$, $s_m$), noise amplitude $U_n$ and $\eta_{th}$, will have the same values of synchronization effectiveness $\eta$ and $SHR_{0,10}$ within one class of figures as a result of neural network symmetry. The initial distribution of all 512 figures into classes allows us to recognize not one specific figure, but a class (out of 102 possible ones) into which this figure is classified. For example, for class $C_5$ the equality $SHR_{0,10}(X_5) = SHR_{0,10}(X_{260}) = SHR_{0,10}(X_{320}) = SHR_{0,10}(X_{65})$ will be observed.

Such classification reduces the number of possible variants to sort as we may send only 102 figures to the input layer, one figure per each class.

The problems this neural network is able to solve may be divided into several variants:

I. Synchronization of oscillators No.0 and No.10 with the corresponding value of $SHR_{0,10}$ and $\eta > \eta_{th}$, exists only for one specific class $C_j$ with number $j = m$ out of 102 classes:

$$SHR_{0,10}(C_j)=\begin{cases} k_0{:}k_{10} \text{ and } \eta \geq \eta_{th} \text{ if } j=m \\ \text{no syncronysation and } \eta<\eta_{th} \text{ if } j \neq m \end{cases} \quad (11)$$

Here we have to show the solutions of this problem with various values of $m$.

II. There is a set of classes $\mathbf{C} = \{C_{Z1}, C_{Z2} \ldots C_{ZP}\}$, where $Z_1, Z_2 \ldots Z_P$ are arbitrary non-repeating indices, where the number is $P < 102$. When inputting this set into the oscillator system, it comes to the synchronization states corresponding to the set $\mathbf{SHR} = \{SHR^{(1)}_{0,10}, SHR^{(2)}_{0,10} \ldots SHR^{(P)}_{0,10}\}$. The set **SHR** does not have the same elements, i.e., each class of figures from set **C** corresponds to a unique synchronization order $SHR_{0,10}$. By analogy with (4) the problem may be expressed as:

$$SHR_{0,10}(C_j)=\begin{cases} SHR^{(1)}_{0,10} \text{ and } \eta \geq \eta_{th} \text{ if } C_j \in \mathbf{C} \text{ and } j=Z_1, \\ SHR^{(2)}_{0,10} \text{ and } \eta \geq \eta_{th} \text{ if } C_j \in \mathbf{C} \text{ and } j=Z_2, \\ \ldots\ldots\ldots\ldots\ldots\ldots\ldots\ldots\ldots\ldots \\ SHR^{(P)}_{0,10} \text{ and } \eta \geq \eta_{th} \text{ if } C_j \in \mathbf{C} \text{ and } j=Z_P, \\ \text{no syncronysation and } \eta<\eta_{th} \text{ if } C_j \notin \mathbf{C} \end{cases} \quad (12)$$

In fact, problem I is a subcase of problem II at $P = 1$.

The principle of solving problem II is shown in Figure 11. Here $P = 3$, and set $\mathbf{C} = \{C_1, C_4, C_5\}$, in this case the corresponding set is $\mathbf{SHR} = \{16{:}15, 13{:}10, 14{:}13\}$. Therefore, unambiguous recognition of figures belonging to three different classes takes place. Synchronization is not realized for all other classes and $\eta < \eta_{th}$.

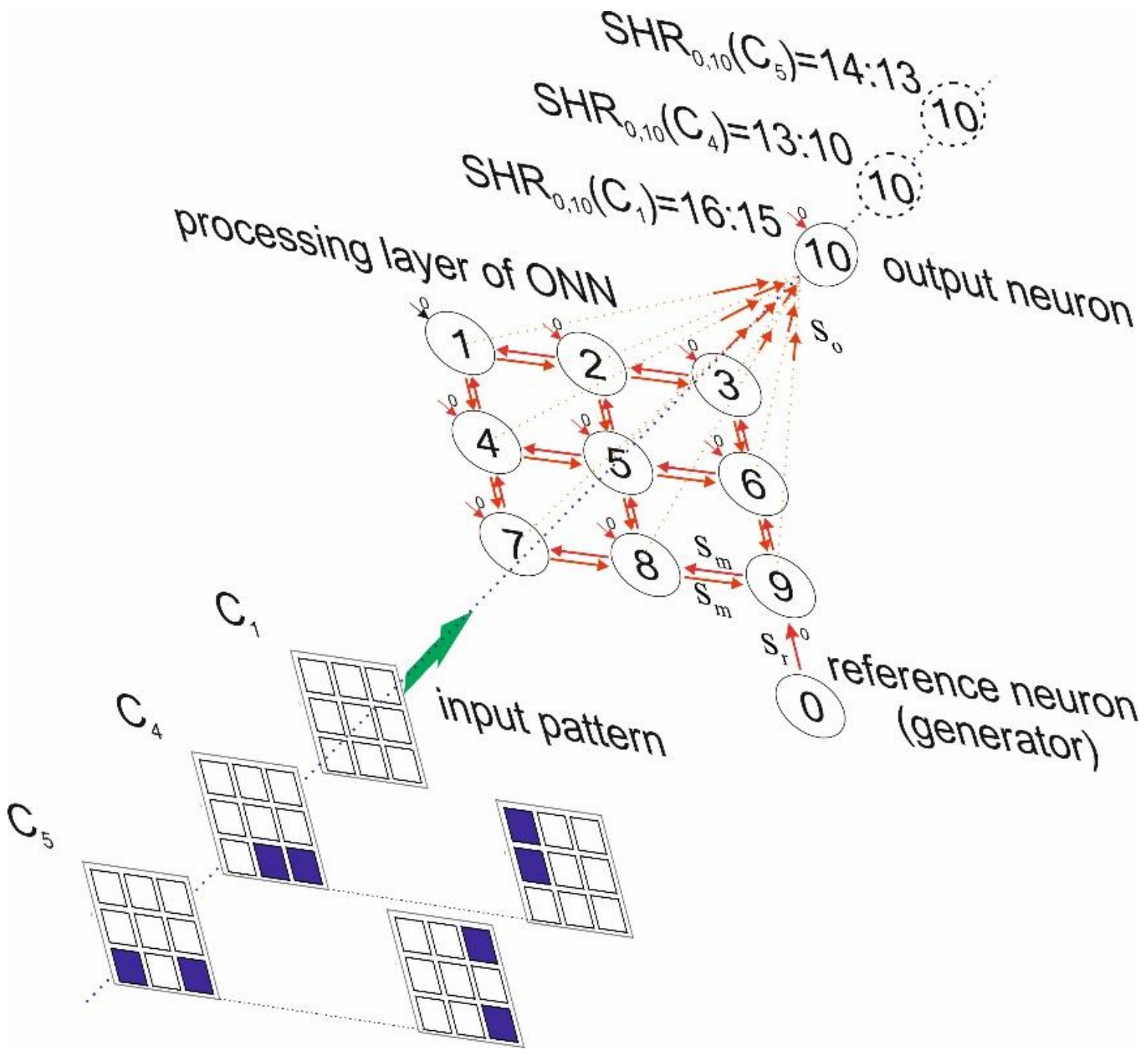

**Figure 11.** The principle of class recognition of neural network figures with one output neuron.

As there is a huge variety of set **C** variants, this problem can be reduced to the search of possible solutions with P > 1 and to the determination of the maximum value of *P*.

III. The third variant of the problem corresponds to a fully trained network when it solves problem II for all possible input classes $C_j$, when $P = 102$.

In this problem, each class out of 102 possible variants corresponds to its unique value of synchronization order $SHR_{0,10}$. Therefore, the set of input classes **C** = $\{C_1, C_2 \ldots C_{102}\}$ transfers into a set of synchronous states **SHR** = $\{SHR^{(1)}_{0,10}, SHR^{(2)}_{0,10} \ldots SHR^{(102)}_{0,10}\}$, where all the elements have non-duplicate values.

Problems I-III have an increasing degree of complexity and are subcases of problem II with different parameter *P*. Problem I is the simplest one and it may be solved by a common neural network with one bistable output neuron (bistability means the presence or absence of synchronization), although two variants of answers in neural networks are often given by using two output neurons [38].

The output neuron in problems II and III have multilevel properties and differ from a common bistable neuron. All three problems can be applied to the circuit shown in Figure 2. The ONN circuit has only one output neuron, nevertheless, the multilevel high order synchronization $SHR_{0,10}$ allows input pattern classification into *P* classes within set **C**. This is the most striking difference between the described neural network and variants presented in the literature. This increases the net data throughput of a single neuron and enables us to create multilevel output cascades of neural networks with high functionality.

## 2.5. Technique for ONN Training

To solve problems I–III, network training is required. As ONN with a high order synchronization effect has not been studied before, there are no established methods for network training. One of the ways is to use a simulated annealing algorithm [39] for the network parameters selection: current ($I_{ON}$, $I_{OFF}$, $I_0$, $I_{10}$), coupling strength ($s_r$, $s_o$, $s_m$), noise amplitude $U_n$ and the synchronization effectiveness threshold $\eta_{th}$. The algorithm's main point is the random searching of the problem II solution with the maximum value of $P$ at some initial interval of parameters, followed by the narrowing of these intervals.

Determination of step size is required for direct parameter searching. In our numerical experiment, the current range was determined by the generation of oscillation in a single oscillator. For the oscillator circuit described in Section 2.1, the generation of oscillation was within the feeding current range of 550 μA ≤ $I_p$ ≤ 1061 μA. Therefore, the currents ($I_{ON}$, $I_{OFF}$, $I_0$, $I_{10}$) varied in this range. We determined the variation steps as $\delta I_{P_i} = 1$ μA. So, there were 512 current steps.

For coupling strength $s$, the variation range was determined by the threshold values of I–V characteristics of a switch. The condition, that is, the integral thermal action on the neuron $s_\Sigma$ should not reduce the threshold voltage of the I–V characteristic of a switch $U_{th}$ below the holding voltage $U_h$, must be met:

$$U_{th} - s_\Sigma > U_h \tag{13}$$

On the basis of Formula (13), for the values of threshold voltages ($U_{th}$ = 5 V, $U_h$ = 1.5 V) and coupling configurations (see Figure 2), the limits of the coupling strength variations are subject to the following conditions: ($s_r$+ $s_o$·9) < 3.5 V and ($s_r$+ $s_m$·4) < 3.5 V. This condition is met with the following ranges that we have chosen: $s_r$ = 0 ÷ 0.2 V, $s_m$ = 0 ÷ 0.5 V, $s_o$ = 0 ÷ 0.3 V. The variation step was chosen as 0.1% of the range magnitude.

The range 20 μV ≤ $U_n$ ≤ 900 μV with the number of steps equal to 12 was chosen for the noise amplitude.

For the synchronization effectiveness threshold $\eta_{th}$, the variation range was 10 % ≤ $\eta_{th}$ < 100%, with the number of steps equal to 25 and minimal spacing $\delta\eta_{th}$ = 0.1%. Parameter $\eta_{th}$ does not belong to the network parameters but rather to the parameters of the algorithm of synchronization order $SHR_{0,10}$ calculation. The value $\eta_{th}$ strongly affects the results of synchronization and the results of pattern recognition because it is a conditionally chosen characteristic. Identifying its optimal value for the recognition problems is an important step in network tuning and training.

Having a lot of network parameters, each parameter's variant should be calculated in 102 classes together with oscillograms and synchronization values $SHR_{0,10}$ at each stage. A full, direct search of all parameters' variants would take a lot of time and computational resources. Therefore, we fixed the values $U_n$ and $\eta_{th}$, and varied currents ($I_{ON}$, $I_{OFF}$, $I_0$, $I_{10}$) and couplings strength ($s_r$, $s_o$, $s_m$). The algorithm for searching for the solution of problem II with the maximum value $P$ included the following steps:

Preparation step: Setting of constant values of $U_n$ and $\eta_{th}$

Step 1: Random searching of parameters ($I_{ON}$, $I_{OFF}$, $I_0$, $I_{10}$, $s_r$, $s_o$, $s_m$) in the maximal range of their variations and finding the values meeting the maximum value $P$. The number of searching attempts is 1000.

Step 2: Narrowing of the parameters ranges by 5 times with their symmetric distribution in relation to the results of the previous step. The number of searching attempts is 1000.

Step 3: Narrowing of the parameters ranges by 25 times with their symmetric distribution in relation to the results of the previous step. The number of searching attempts is 1000.

Noise amplitude $U_n$ is kept fixed because this is the parameter that is not often controlled in the experiment. It is determined by external and internal circuit noises, and we considered its effect individually in more detail. We used $U_n$ = 80 μV, which is close to the experimentally observed value [19].

The threshold value of synchronization effectiveness $\eta_{th}$ is fixed because it is the main parameter in the synchronization algorithm, and we considered its effect individually in more detail. For the pattern recognition experiments, we used a workstation (Intel Xeon quad core processor, Albuquerque , New Mexico, USA, 4 × 2 GHz, 8GB RAM) running 64-bit Windows Server 2008. CPU time for a single run on one core for the direct search procedure with 1000 search attempts took ~5 h.

## 3. Results

Following the proposed algorithm, we fixed the values of the following parameters ($U_n$ = 80 µV, $\eta_{th}$ = 90%) and varied the currents ($I_{ON}$, $I_{OFF}$, $I_0$, $I_{10}$) and coupling strength ($s_r$, $s_o$, $s_m$). The distribution of the solutions after the first, second and third steps of training is shown in the diagram (Figure 12), the values of *P* are on the x-axis, and the corresponding number of solutions $N_P$ are on the y-axis. The largest number of solutions corresponds to *P* = 0 when there is no synchronization at any input class $C_j$. After steps 2–3. the number of solutions with a low value of *P* decreases and the solutions with a higher value of *P* appear. Step 3 gives more solutions in the range *P* = 6–10 in comparison with step 2, although the maximum value *P* = 14 is similar in both cases.

The incorrect solution of training (see Figure 13a), which does not correspond to the problem I-III conditions, occurs rather often. For example, two or more classes may correspond to the same value of $SHR_{0,10}$, and the probability of such an event at the first step of training is ~55% (the calculation was based on the data in Figure 12), which results in the ambiguous recognition of figures and their classes. Another frequent case of wrong training is the absence of oscillatory synchronization for any input class (*P* = 0), which has a probability of ~30% in the first step (see the values in Figure 12).

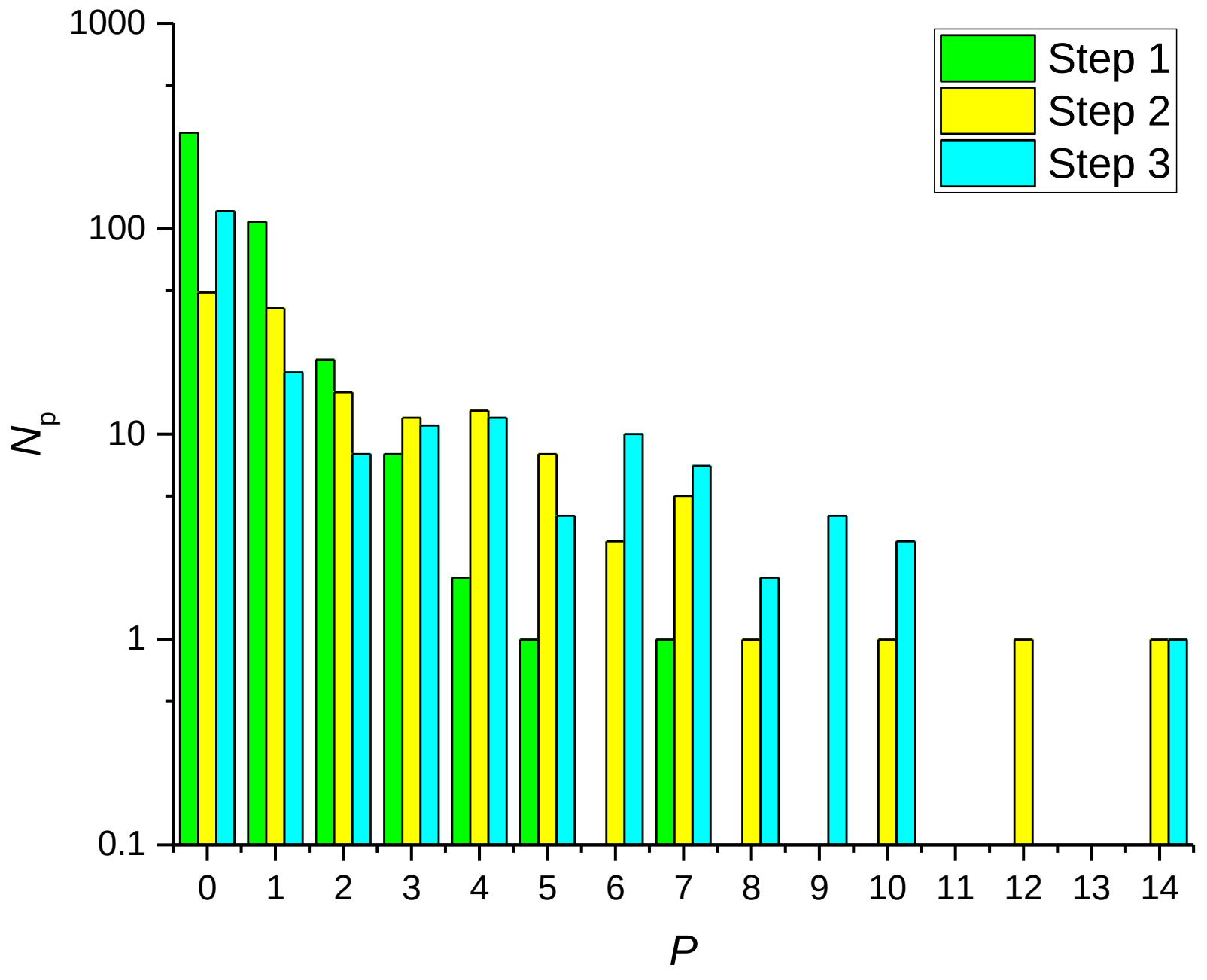

**Figure 12.** Diagram of the solution number $N_P$ distribution against the value of *P* for three subsequent steps of the network training algorithm. The total number of attempts at each step is 1000. The maximum value *P* = 14 was obtained at the following parameters: ($I_{ON}$ = 722 µA, $I_{OFF}$ = 1034 µA, $I_0$ = 1020 µA, $I_{10}$ = 887 µA, $s_r$ =0.10176 V, $s_o$ = 0.29202 V, $s_m$ = 0.202 V, $U_n$ = 80 µV, $\eta_{th}$ = 90%).

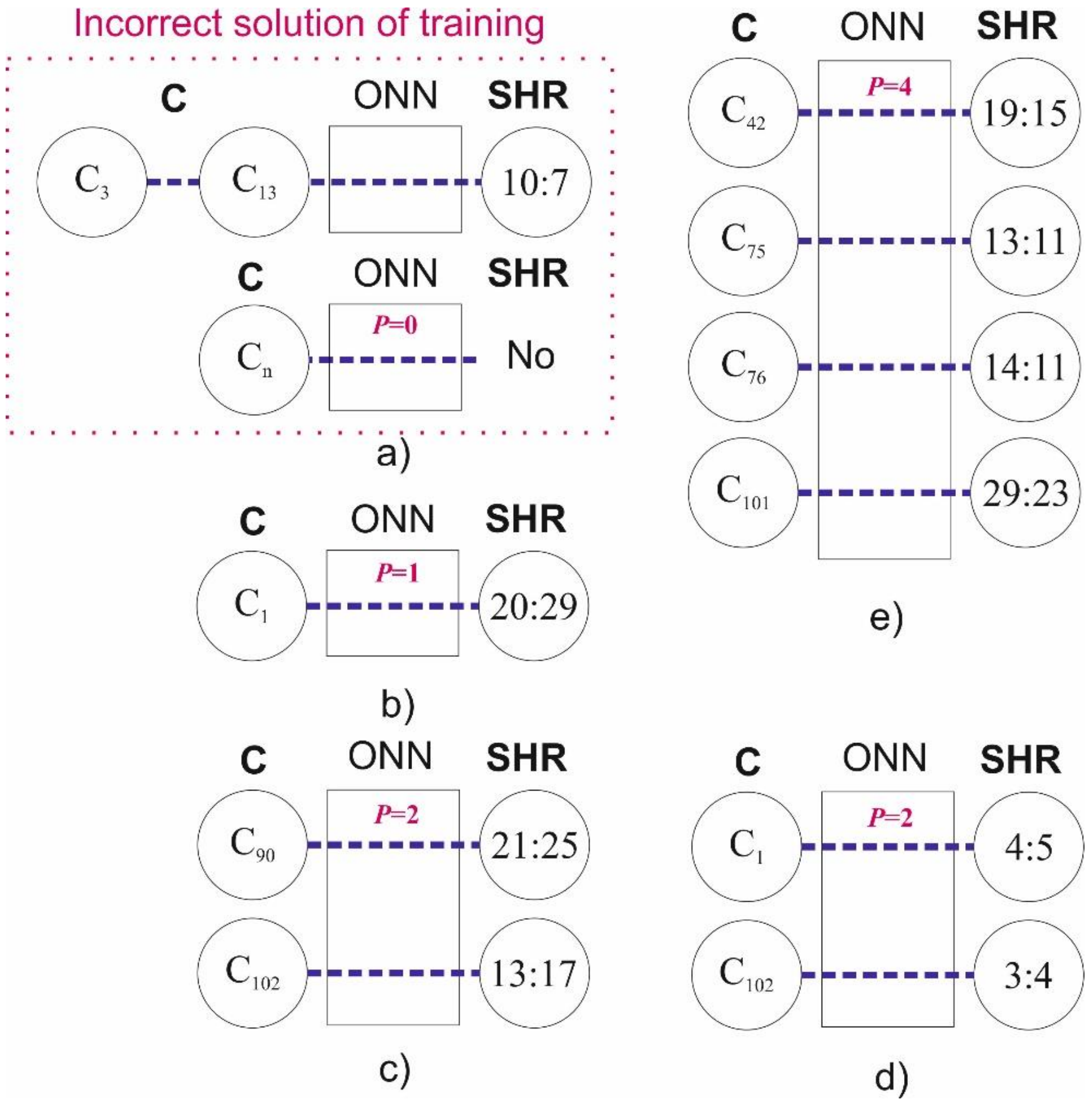

**Figure 13.** Training results at various values of current $I_{ON}$: examples of incorrect solutions of training (**a**) and solutions of problem I (**b**); examples of correct solutions of problem II (**c**–**e**). The complete list of the model parameters is presented in Supplementary Materials (Data2.txt).

### *3.1. Solution of Problem I*

The experiment demonstrated that the solution for problem I ($P$ = 1) is easily found for some classes $C_m$. For example, Figure 13b shows a neural network that recognizes class $C_1$ with the corresponding synchronization order $SHR_{0,10}$ = 20:29, and there is no synchronization for all other classes. The probability of any solution with $P$ = 1 during the first step is ~10% (here the total number of attempts is 1000 and the value of $N_P$ ~ 100, see Figure 12), and this is the maximum probability of the solution in comparison with other $P$ > 1. Figure 14 shows the distribution of solutions at $P$ = 1 for various values of $m$ (4) after 60 repetitions of the first step of training at the maximum range of all parameter variations. The maximum probability (~4%) can be seen for solutions with sets $C_1$ and $C_{102}$ when all cells of the input pattern are either empty or colored, see Figure 9. Solutions for other $m$ are much rarer, with the probability being two orders lower (~0.03%). Parameters corresponding to the same solution can differ significantly. For example, the system can recognize class $C_{102}$ at the input, while at the output $SHR_{0,10}$ would be different for different parameters, however, the problem is still considered solved. The histogram shows that the network can be trained to solve problem I with a certain, predetermined value of $m$.

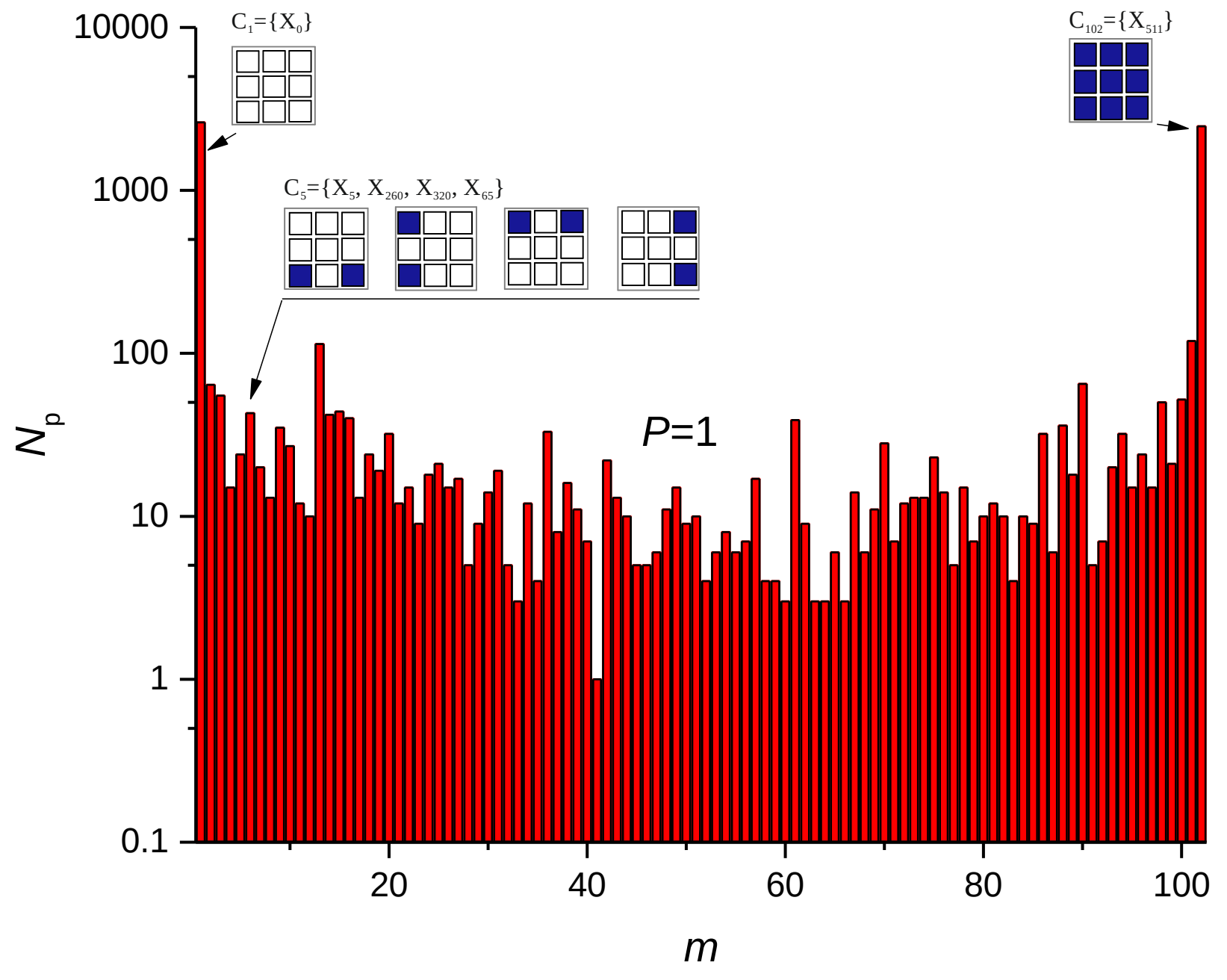

**Figure 14.** Distribution of the solutions number for problem I ($P$ = 1) according to the values $m$ (4), calculated for 60 repetitions of the first step of training ($U_n$ = 80 μV, $\eta_{th}$=90%). Insets show the classes of input images $C_m$ for m = 1, m = 5, m = 102.

## *3.2. Solution of Problem II*

The condition of problem II requires that a certain class from set **C** should correspond to the unique synchronization order from set **SHR** while there is no output oscillator synchronization for other classes. For example, at the first step of training, $N_P$ = 26 solutions were found for $P$ = 2, two of them are shown in Figure 13c,d, and Figure 13e demonstrates the variant of set **C** and **SHR** for $P$ = 4.

We can find solutions with a random search, for example, for $P$ = 2, however, these solutions would be for different satisfying condition (12). For $P$ = 2, there are $C^2_{102}$ = 5151 solution combinations (where $C^2_{102}$ is the number of combinations 2 out of 102). For $P$ = 14, $C^{14}_{102}$ is a 17-digit number. Therefore, we indicate the maximum value of $P$, and do not indicate which solution was found.

After all training steps, the maximum value of $P$ reached $P$ = 14. While ONN configuration and training algorithm can be further improved, the purpose of this work is to demonstrate that a multilevel neural network can be implemented and used for pattern recognition.

## *3.3. Solution of Problem III*

The solution for problem III has not been found yet (see Section 4).

## *3.4. Study of the Noise Effect on the Training Results*

We constructed a 3D graph as shown in Figure 15, to find the dependence of maximum possible $P$ and $N_P$ on the noise amplitude value in network $U_n$ ($\eta_{th}$ = 90%). The values were taken from the first step of training.

When the noise increases above $U_n$ = 400 μV, the number of solutions $N_P$ and the value $P$ sharply decreases. Most of the solution variants are distributed in the range 1 ≤ $P$ ≤ 2. The maximum value for $N_P$ corresponds to $P$ = 1 at noise level $U_n$ = 400 μV. In this case, the integral value $\sum N_P$ for all values $P$ also has a maximum for $U_n$ = 400 μV. Therefore, "stochastic resonance" is present when a certain level of noise induces maximum number of solutions for problems I–II. This may be caused by two differently directed tendencies of $N_P$ reduction: the occurrence of "extra" synchronizations with decreasing $U_n$ and suppression of the number of synchronizations with increasing $U_n$.

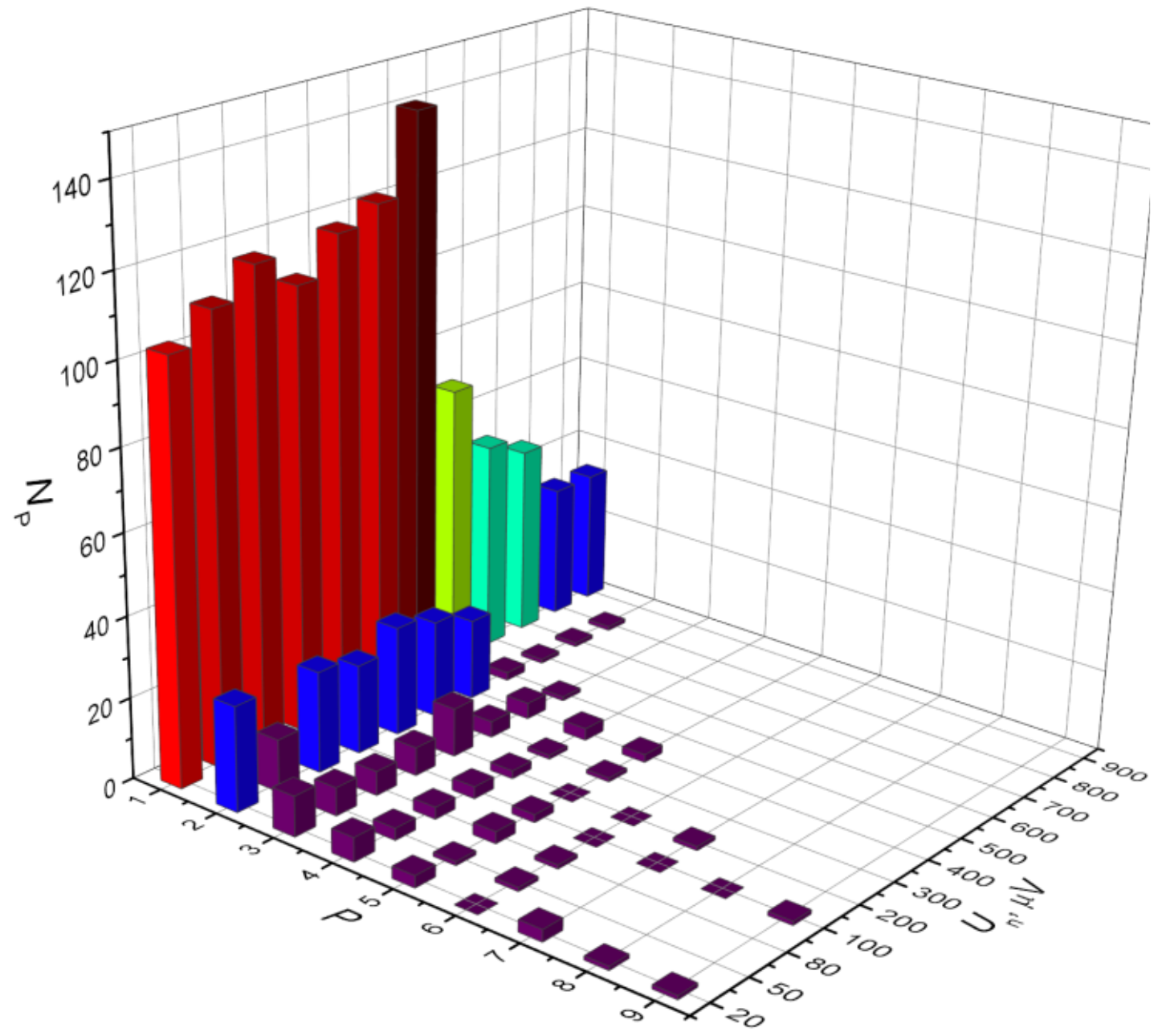

**Figure 15.** Dependence of solution number $N_P$ for various values of $P$ on noise level $U_n$ ($\eta_{th}$ = 90%).

### *3.5. Examination of the Synchronization Threshold on the Training Result*

We constructed a 3D graph as shown in Figure 16, to find the dependence of maximum possible $P$ and $N_P$ on the value of threshold synchronization effectiveness ($U_n$ = 80 µV).

A general tendency for reduction of $N_P$ with a decrease in $\eta_{th}$ below 40% can be seen, and the growth of $\eta_{th}$ up to $\eta_{th}$ = 99% on average does not change the values of $P$ and $N_P$. This seems to be caused by the fact that synchronization occurring during problems I and II solution has a high value of effectiveness $\eta$ > 99%. In this case, reduction of $\eta_{th}$ just adds "extra" synchronous states, which do not meet the problem conditions.

The fact that the maximum possible $P$ reached $P$ = 11 at $\eta_{th}$ = 30% is interesting, and we observed some resonance for values of $P$. Maximum $P$ declines as with reduction of $\eta_{th}$ as with growth of $\eta_{th}$, caused by reduction of in the solution number $N_P$.

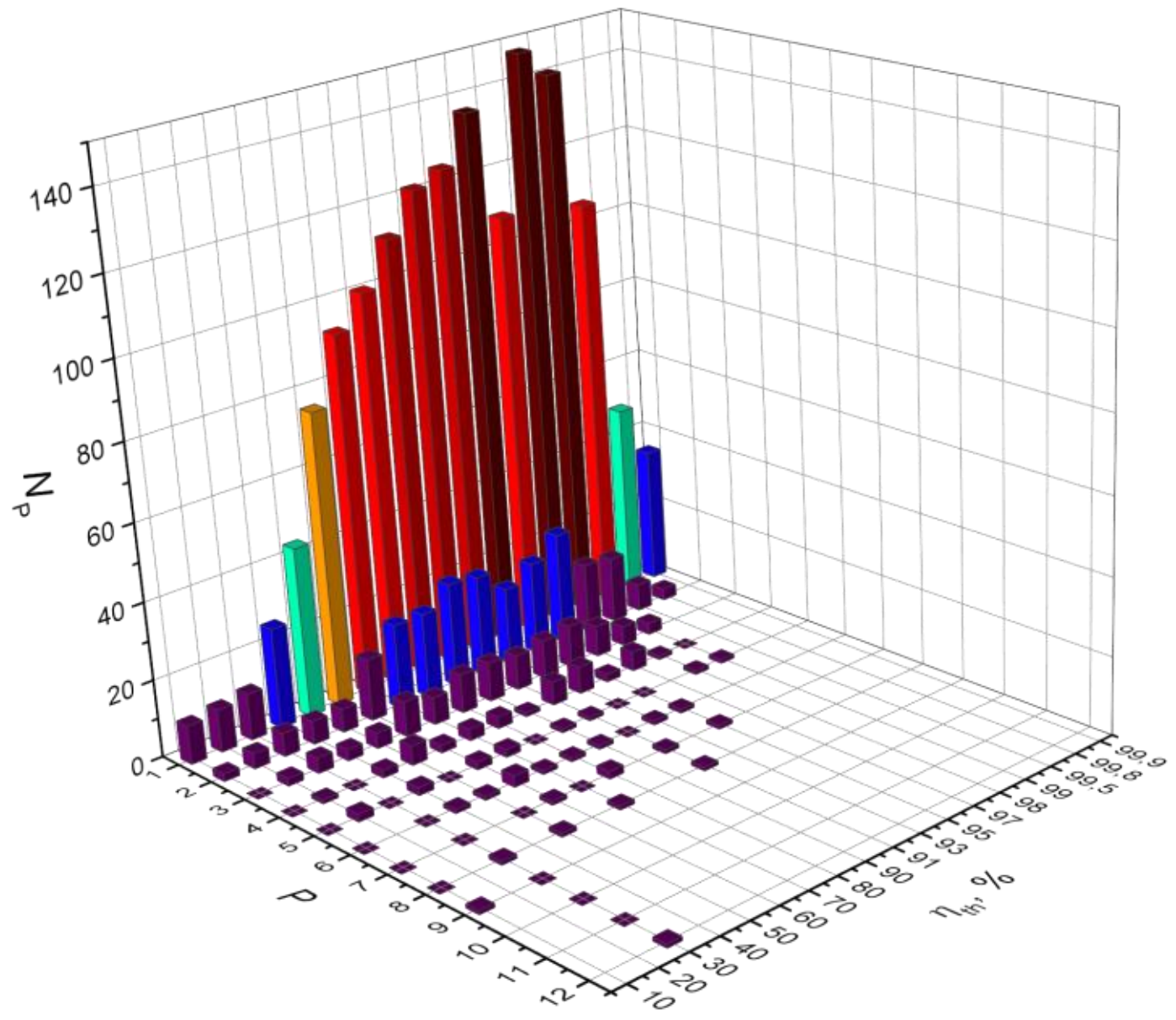

**Figure 16.** Dependence of solution number for various values of $P$ on the threshold synchronization effectiveness $\eta_{th}$ ($U_n$ = 80 µV).

### *3.6. Study of the Dynamics of the Neural Network*

To understand how the output oscillator changes its synchronization relative to the reference oscillator, and what role the layer of processing neurons plays in the process, we conducted the following model experiments.

First, we turned off the effect of input oscillators on the output layer, making $s_o$ = 0 V. The result was a circuit equivalent to two coupled oscillators, when oscillator No.0 affects oscillator No.10 with a force of $s_r$ = 0.3 V. By varying the oscillator currents ($I_0$, $I_{10}$), we obtained the standard distribution pattern of $SHR_{0,10}$ in the form of Arnold tongue, where the regions of equal synchronization are extended sections in the form of rays (see Figure 17a) with diagonal symmetry. A special feature is a smooth (gradient) increase in $SHR_{0,10}$ from the lower right distribution corner ($I_0$ = 1061 μA, $I_{10}$ = 500 μA) to the upper left corner ($I_0$ = 500 μA, $I_{10}$ = 1061 μA). Similar patterns of synchronization distribution in the system of two oscillators were observed by us earlier, for example, in [33]. After adding the effect of the processing layer, when $s_o$ = 0.1 V, we get the type of synchronization distribution shown in Figure 17b. The synchronization areas are island type. The layer of input oscillators divides Arnold's tongue into local areas while the tendency of the gradient increase in synchronization is preserved.

If we fix the currents $I_0$ and $I_{10}$, when $s_o$ = 0.29 V, and vary the currents $I_{ON}$ and $I_{OFF}$, the pattern of the synchronization distribution changes its appearance (see Figure 18a). We observed a chaotic scatter of the synchronization magnitude over the field of parameters, with preservation of the diagonal symmetry and a wide scatter of values within 1–200 μA. As $I_{ON}$ and $I_{OFF}$ values change the frequency of many oscillators in the processing layer at once, it is difficult to predict the tendency of the synchronization distribution. Nevertheless, the range of $SHR_{0,10}$ variations is much smaller than when $I_0$ and $I_{10}$ are varied, which is obviously due to the constancy of the main frequency between the oscillators whose synchronization is measured. With a decrease in the value of $s_o$ to $s_o$ = 0.1 V, the form of the distribution has a similar appearance, but the range of $SHR_{0,10}$ change is reduced.

In analyzing Figures 17 and 18, we can assume that the model annealing method, where we randomly select the system parameters, is searching for optimal combinations of synchronization regions. Due to the large number of regions and their unpredictable order, it is possible to find a solution to problems I–II, and it is not unique, as we observed in previous experiments.

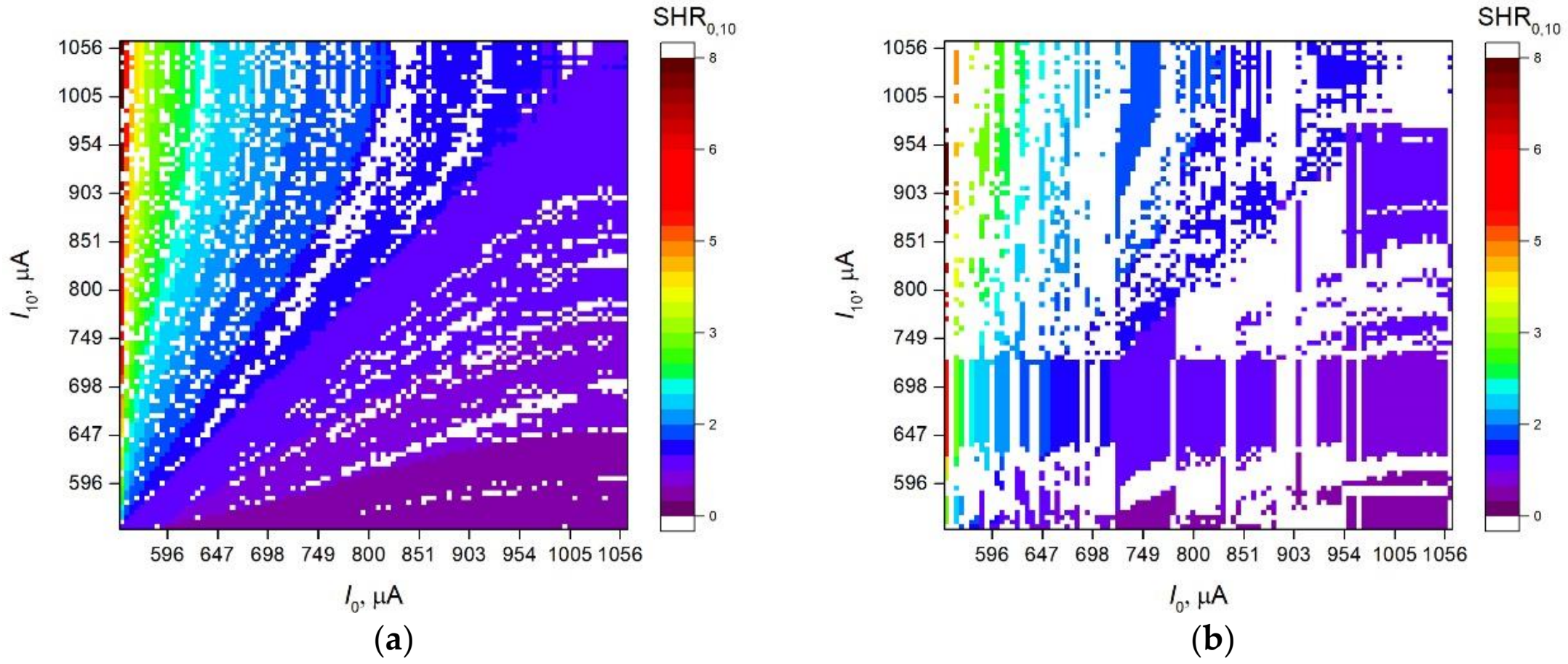

**Figure 17.** The synchronization distribution $SHR_{0,10}$ in the regions of currents $I_0$ and $I_{10}$ with (**a**) $s_o$ = 0 V and (**b**) $s_o$ = 0.1 V. For all cases, $I_{ON}$ = 725 μA, $I_{OFF}$ = 1036 μA, $s_r$ = 0.3 V, $s_m$ = 0.207 V, $\eta_{th}$ =90%, $U_n$ = 80 μV, and the class $C_{94}$ is the input.

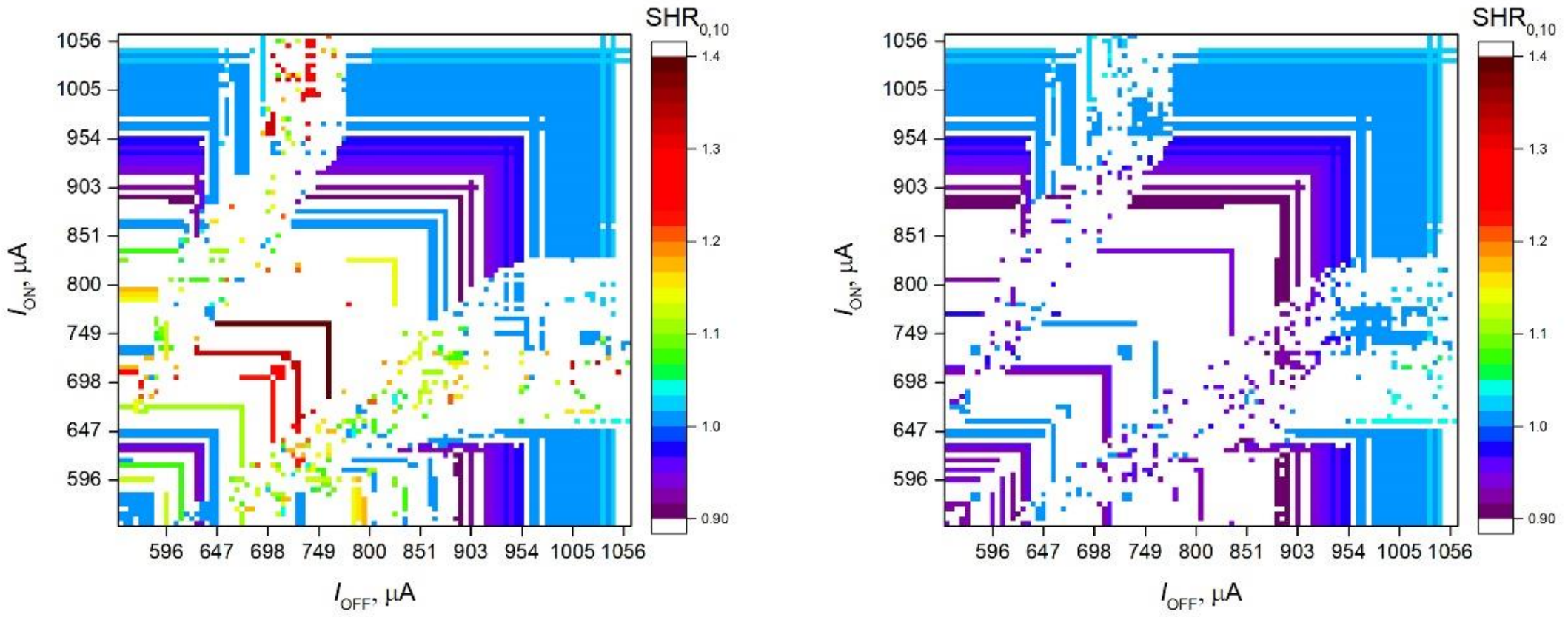

(**a**) (**b**)

**Figure 18.** The synchronization distribution $SHR_{0,10}$ in the regions of currents $I_{ON}$ and $I_{OFF}$ with (**a**) $s_o$ = 0.29V and (**b**) $s_o$ = 0.1 V. For all cases, $I_0$ = 1017 μA, $I_{10}$ = 891 μA, $s_r$ = 0.1036 V, $s_m$ = 0.207 V, $\eta_{th}$ = 90%, $U_n$ = 80 μV, and the class $C_{94}$ is on the input.

The following numerical experiment shows some features of the network operation dynamics. The idea was to find out how the matrix of the processing layer affects the synchronization of oscillators No.0 and No.10. Let us consider two cases. In the first case, $s_o$ = 0 V, $s_r$ = 0.13 V, $I_0$ = 650 μA, $I_{10}$ = 950 μA, there is only one unidirectional effect of oscillator No.0 on No.10, and the base synchronization has the value $SHR^b_{0,10}$ = 23:12 = 1.917. In the second case, we added random variations of coupling strength ($s_o$, $s_m$) and currents ($I_{ON}$, $I_{OFF}$) at the same values of $s_r$ = 0.13, $I_0$ = 650 μA, $I_{10}$ = 950 μA. As a result, after the first step of training we obtained a set of solutions (**C** and **SHR**) for various P > 0 and positioned all elements of $SHR_{0,10}$ on the plot (see Figure 19). It can be seen that all elements of the sets **SHR** are distributed within some interval above the boundary $SHR^b_{0,10}$. The dispersion is $\delta SHR_{0,10}$ ~0.6.

The same behavior is observed when the current values were interchanged ($I_0$ = 950 μA and $I_{10}$ = 650 μA). The base synchronization had the value of $SHR^b_{0,10}$ = 8:15 = 0.533, and when other parameters were varied, the elements of **SHR** diverge upward with $\delta SHR_{0,10}$ ~ 0.4. At high and similar values of current $I_0$ = 1058 μA and $I_{10}$ = 1058 μA, the base synchronization is $SHR^b_{0,10}$ = 1:1 = 1, and $\delta SHR_{0,10}$ ~ 0.3.

Thus, the matrix of processing layer oscillators only deviates the $SHR_{0,10}$ value upward from the base synchronization value of oscillators No.0 and No.10. The effect of oscillators positioned in the processing layer only increases the frequency of oscillator No.10, but cannot decrease it because of the nature of thermal coupling, which can initiate switching but cannot suppress it.

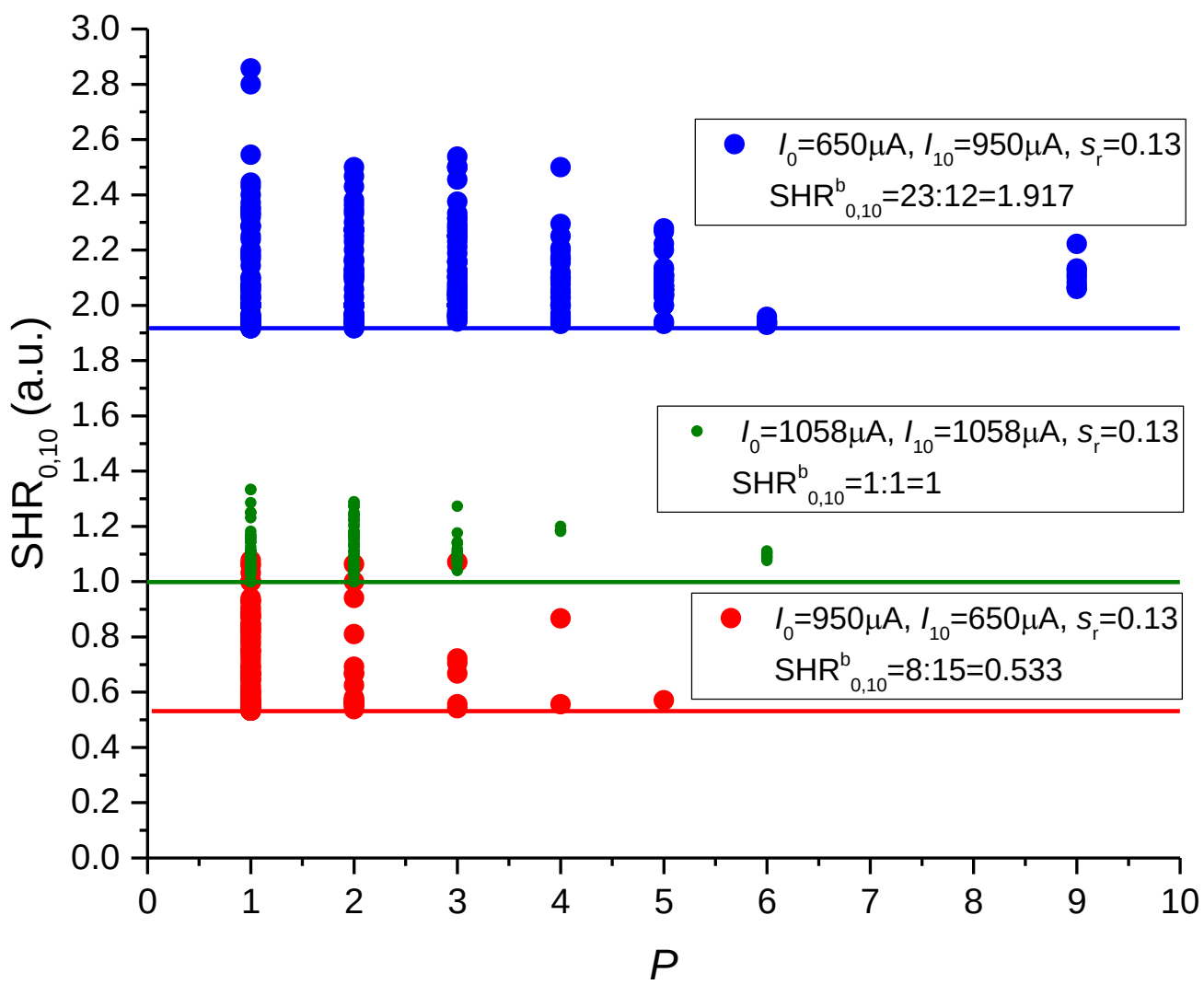

**Figure 19.** Distribution of synchronization values $SHR_{0,10}$ at random variations of coupling strengths ($s_0$, $s_m$) and currents ($I_{ON}$, $I_{OFF}$). Fixed parameters are shown in the plot. The base sync value ($SHR^b_{0,10}$) is shown by a solid line. $SHR^b_{0,10}$ is the sync value at $s_0 = 0$.

## 4. Discussion

In the previous section, we demonstrated how the simulated annealing algorithm can be used for ONN training. In this algorithm, we used three steps for parameter searching, however, we did not manage to surpass the maximum value of $P = 14$ (see Figure 12). Therefore, we can assume that the simulated annealing method reaches a solution limit, and to search the network parameters with higher $P$, it might be necessary to vary other ONN properties, for example, the number of oscillators of the processing layer, the I–V characteristics of oscillators, and the matrix of couplings $S$. Furthermore, the values of $U_n$ could be selected more carefully on the basis of the regularities shown in Figure 15, where the maximum of the solution number $N_P$ at a certain noise level is presented. This effect is similar to stochastic resonance phenomena in spike networks [40–42], and requires additional research and analysis.

The simulated annealing algorithm might not be an effective training method, and development of more efficient algorithms for this type of network is a separate global problem for future research.

For proper network training, a deeper understanding of the physics of the oscillators' synchronization process is necessary. For example, the results and regularities arising in Figure 19 assist in understanding the process of solution generation. We demonstrated the presence of a boundary for SHR values that is determined by the base synchronization value $SHR^b_{0,10}$ of oscillators No.0 and No.10 and showed that the effect of oscillators' processing layer leads to $SHR_{0,10} \geq SHR^b_0$. The selection of currents for the reference oscillators No.0 and No.10 determines the variation range of $\delta SHR_{0,10}$, and most likely, the larger this range is, the higher the probability of finding the solutions with high $P$. If we set the current values $I_0$, $I_{10}$, for example, at the boundary of the maximum range $I_0 = 1058$ μA, $I_{10} = 1058$ μA, then it will be more difficult for the matrix of oscillators in the processing layer to change the dynamics of the output oscillator that operates in a high frequency mode and under the high frequency effect of the reference oscillator. As a result, we observed narrowing of $\delta SHR_{0,10}$ to ~0.3 (see Figure 19). Figure 18 demonstrates the areas of parameters where $SHR_{0,10}$ practically does not change and where it changes very often, therefore, this may explain the productivity of the annealing training algorithm. In Step 1, we found areas with a large number of solutions, and then in Step 2 and 3, by thorough scanning of these areas, we found solutions with the largest $P$ (see Figure 12).

In addition to the configuration and network parameters, the selection of parameter $\eta_{th}$, which is used in the synchronization magnitude determination method, is of great importance. As the results and regularities arising in Figure 16 show, too high ($\eta_{th} \geq 99.8$) and too low ($\eta_{th} \leq 20$) parameters significantly reduce the number of solutions $N_P$. Nevertheless, we suppose that the usage of $\eta_{th} < 50\%$ is not completely justified, as the parameter $\eta$ determines the share of a synchronous signal in the total oscillogram [19]. Moreover, the technique of SHR determination may affect the result, and future research aimed at its improvement may lead to positive shifts in this field. The addition of alternative methods of pattern transfer into a network that vary, for example, current $I_{10}$ or couplings magnitudes may significantly expand the range of variations of $SHR_{0,10}$, hence, the maximal attainable value $P$.

The network structure defined by the coupling matrix $S$ (1) was chosen in order to implement an actual $VO_2$–based device, therefore, oscillator No.0 affected all other oscillators. This can be easily implemented by arranging $VO_2$-switches on one substrate. Nevertheless, the development of an actual device requires specific setting of coupling strengths that depend on many parameters [19] and this is beyond the scope of the current work.

Although problem III has not been solved, the presented results and solution for problem II for $P = 14$ showed that multilevel high order synchronization increases net data throughput of a single neuron and enables the creation of multilevel output cascades of neural networks with high functionality.

The duration of the processed oscillograms was ~2.5 s (for the number of pulses ~3000), providing an error less than 0.2% for calculating $\eta$ (see Appendix A.2). In the experiment, such duration would significantly slow down the work with the system, since the time of the synchronization calculation would be several seconds. Nevertheless, we hope for a real implementation of this idea on oscillators with a large generation frequency (~ 1–10 MHz), which would significantly reduce the oscillogram duration and the time for the synchronization calculation.

## 5. Conclusions

The paper presents a new model of ONN with high order harmonics synchronization that recognizes and classifies visual patterns by unique synchronous states, i.e., in accordance with their definite symmetry class. The most outstanding feature of this neural network, in comparison with published variants, is that neurons possess not only bistable properties (the presence or absence of synchronization with the reference neuron) but exhibit multilevel synchronization. The presented model has only one output neuron, nevertheless, variation in the value of high order synchronization $SHR_{0,10}$ allows its usage to classify the input pattern into $P$ classes with the set **C**. The training implementation of this network for solving cognitive tasks is an interesting possibility, as the network consists only of oscillators and does not use other computational modules.

The main purpose of the article was to demonstrate a new method of pattern recognition based on coupled oscillators, where cognitive functions are realized due to the high order synchronization effect. This is a universal effect, so regardless of the type of connection, thermal or any other, the cognitive functions can be realized using coupled oscillators. Thermal communication has no special role in pattern recognition, it is only necessary for organizing the synchronization of oscillators.

The development of ONN models with high order synchronization effect offers significant potential for increasing the efficiency of artificial intelligence networks, and the development of their training techniques is an important direction for further research. The possibility of implementing this ONN as not only a program code but also as a separately operating device based on real micro- and nano-electronic self-oscillators would enhance the importance of the obtained results and promote further research in this field.

**Supplementary Materials:** The following are available online at list of classes: Data1.txt, Data2.txt.

**Author Contributions:** Conceptualization, A.V.; methodology, A.V., M.B. and P.B.; software, A.V.; validation, P.B.; writing—original draft preparation, A.V., M.B and P.B.; project administration, A.V.

**Funding:** This research was supported by the Russian Science Foundation (grant no. 16-19-00135).

**Acknowledgments:** The authors express their gratitude to O. Dobrynina and Andrei Rikkiev for the valuable comments in the course of the article translation and revision.

**Conflicts of Interest:** The authors declare no conflict of interest.

## Appendix A

### *Appendix A.1. Model Circuit of a Coupled Oscillators-Based Neural Network*

The circuit of the modelled ONN is presented in Figure 2. It consists of 11 oscillators numbered $i$ = 0..10. The magnitude of the $i$-th oscillator effect on the $j$-th oscillator is determined by the coupling strength $s_{i,j}$, which is set by matrix $S$ (1).

The model circuit of a single oscillator is given in Figure A1. It includes a source of current $I_{P_i}$, capacitance $C$ connected in parallel with a $VO_2$ switch and noise source $U_{n_i}$. The capacity magnitude $C$ was constant $C$ = 1 00 nF, but $I_{P_i}$ and $U_{n_i}$ varied within the following ranges: $I_P$ (550 μA ÷ 1061 μA), $U_n$ (20 μV ÷ 900 μV). The current through the $VO_2$ switch and its voltage are defined as $I_{sw_i}$ and $U_i$, respectively. Voltage-current characteristics of the $VO_2$ switch are shown in Figure A2, where the experimental and model curves are presented.

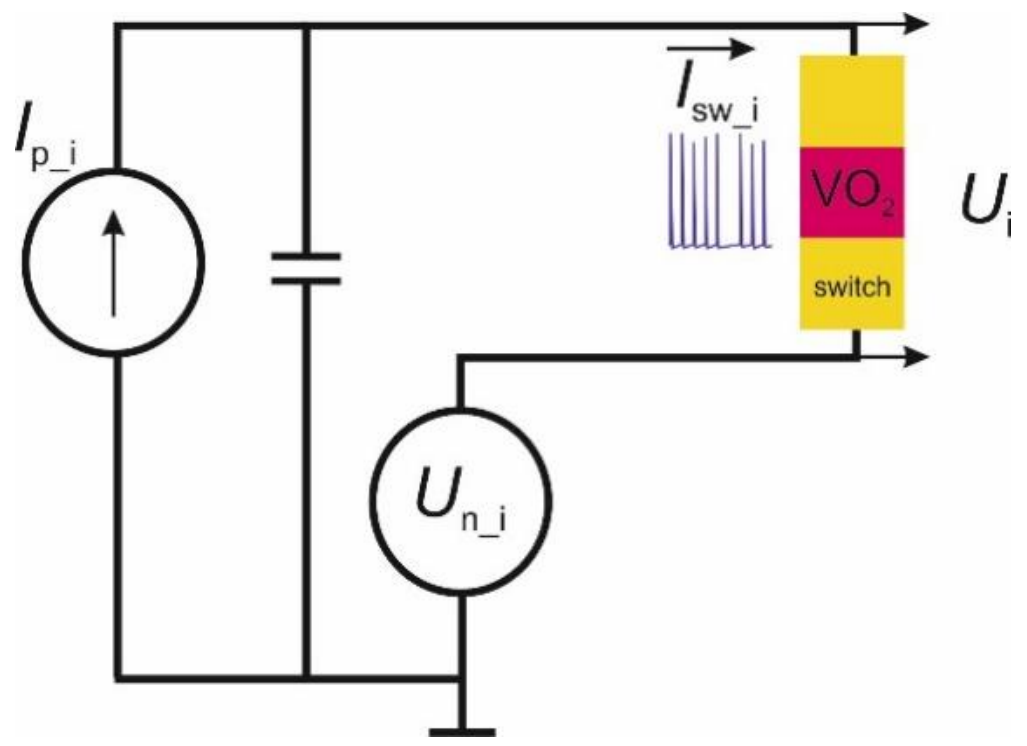

**Figure A1.** Circuit of a single oscillator of a $VO_2$-based structure. *I* is the number of an oscillator; $I_{p_i}$ is a source of current, *C*is a capacitance, $U_{n_i}$ is a source of noise, $I_{sw_i}$ is the current through the $VO_2$ switch, $U_i$ is the voltage at the switch.

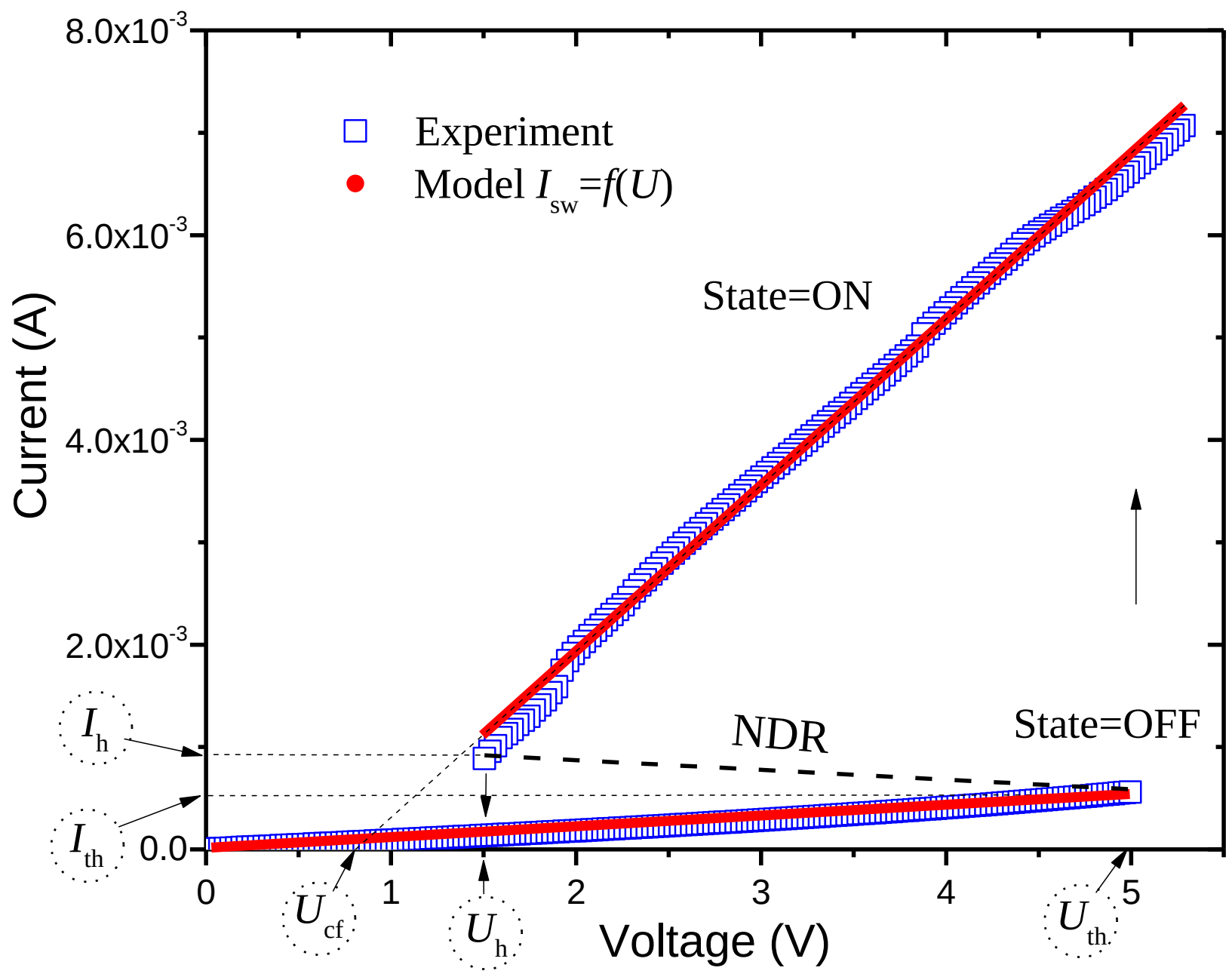

**Figure A2.** Typical experimental I–V characteristics of a separate switch (Experiment curve) and its model curve (Model curve).

All model switches without coupling have the same I–V characteristics, with stationary natural parameters $U_{th}$ = 5 V (threshold voltage), $U_h$ = 1.5 V (holder voltage), $U_{cf}$ = 0.82 V (cutoff voltage).

The model curve $I_{sw_i}$ = $f(U_i)$, which has high-resistance (OFF) and low-resistance (ON) segments with corresponding dynamic resistances $R_{off_i}$ = 9100 Ω and $R_{on_i}$ = 615 Ω, can be presented by the following formula:

$$I_{sw_i} = f_i(U) = \begin{cases} \dfrac{U_i}{R_{off_i}}, & \text{if } flag_i = 1 \text{ (OFF)} \\ \dfrac{(U_i - U_{cf_i})}{R_{on_i}}, & \text{if } flag_i = 0 \text{ (ON)} \end{cases} \quad \text{(A1)}$$

where $i$ = 0 … 10 is the oscillator number, $flag_i$ denotes the $VO_2$ switch state 1 (OFF—high resistance), 0 (ON—low resistance).

Transitions from one state into another can be expressed by the following algorithm:

$$flag_i = \begin{cases} 1 \text{ (OFF)}, & \text{if } (flag_{\mathrm{i}} = 0) \text{ and } (U_{\mathrm{i}} < U_{\mathrm{h_i}}) \\ 0 \text{ (ON)}, & \text{if } (flag_{\mathrm{i}} = 1) \text{ and } (U_{\mathrm{i}} > U_{\mathrm{th_i}}) \end{cases} \quad \text{(A2)}$$

where $U_{\mathrm{th_i}}$ and $U_{\mathrm{h_i}}$ are threshold turn-on and holder voltages of switches (see Figure A3):

As mentioned in Section 2.1., the model of thermal coupling is based on reduction of threshold voltage $U_{\mathrm{th}}$ by the value *s* due to thermal effect of other switches. The value *s* is the thermal coupling strength. The physics of thermal interaction is caused by the generation of a heat wave when the switch is turned on, which propagates and acts (heats) on the surrounding switching structures. In [19], we showed that it is possible to introduce an interaction radius $R_{\mathrm{TC}}$ beyond which the induced temperature is less than 0.2 K. Therefore, in the model, we assume that each switch only interacts with surrounding switches that are within the radius of $R_{\mathrm{TC}}$. The heating of structures leads to a decrease in the threshold voltage [19], this change characterizes the value of *s*, which we call the coupling force. The higher the *s*, the stronger the effect of one switch on the other.

Coupling of oscillators results in changing their threshold turn-on voltages $U_{\mathrm{th_i}}$ in regard to the switch state ($flag_{\mathrm{i}}$) with which they interact.

In general, the magnitude $U_{\mathrm{th\,i}}$ can be presented by the following formula:

$$U_{\mathrm{th_i}} = U_{\mathrm{th}} - \sum_{j(\text{if } flag_{\mathrm{j}}=0 \text{ (ON)})} s_{\mathrm{j,i}} \quad \text{(A3)}$$

where *j* runs though all values, at which the switch is in turn-on state ($flag_{\mathrm{i}} = 0$), and $U_{\mathrm{th}}$ is the natural turn-on voltage without coupling.

This means that all effects of $s_{\mathrm{j,i}}$ on *i*-switch are summed up from all the other *j*-th switches at the turn-on state.

The differential equation that describes the circuit operation (Figure A2) is written as:

$$C\frac{dU_{\mathrm{c_i}}(t)}{dt} = I_{\mathrm{p_i}} - I_{\mathrm{sw_i}}(t), \quad \text{were } I_{\mathrm{sw_i}}(t) = f(U_{\mathrm{c_i}}(t) - U_{\mathrm{n_i}}(t)) \quad \text{(A4)}$$

where *I* is the number of an oscillator, *C* is capacitance, $U_{\mathrm{c_i}}$ is the capacitor voltage, $I_{\mathrm{sw_i}}$ is the current through the switch, $U_{\mathrm{n_i}}$ is the noise in the switch, and *f*(*U*) is the I–V characteristic function (A1).

Equations (A4) were numerically calculated with respect to time at regular intervals $\Delta t = 10^{-5}$ s using implicit Euler method and discrete noise $U_{\mathrm{n}}(t)$ was generated according to the algorithm $U_{\mathrm{n}}(t) = U_{\mathrm{n0}} \cdot \mathrm{randn}(t)$, where $U_{\mathrm{n0}}$—noise amplitude and randn(*t*)—normal random numbers generated with zero mean and dispersion equal to 1, realized through the algorithm of uniformly distributed random value transformation [43].

As a result, by setting oscillator current values $I_{\mathrm{p_i}}$, noise amplitude $U_{\mathrm{n0}}$, and coupling strengths ($s_{\mathrm{r}}$, $s_{\mathrm{m}}$, $s_{\mathrm{o}}$) we could calculate oscillograms of current $I_{\mathrm{sw_i}}(t)$, voltage at the capacity $U_{\mathrm{c_i}}(t)$ and voltage at switches:

$$U_{\mathrm{i}}(t) = U_{\mathrm{c_i}}(t) - U_{\mathrm{n_i}}(t) \quad \text{(A5)}$$

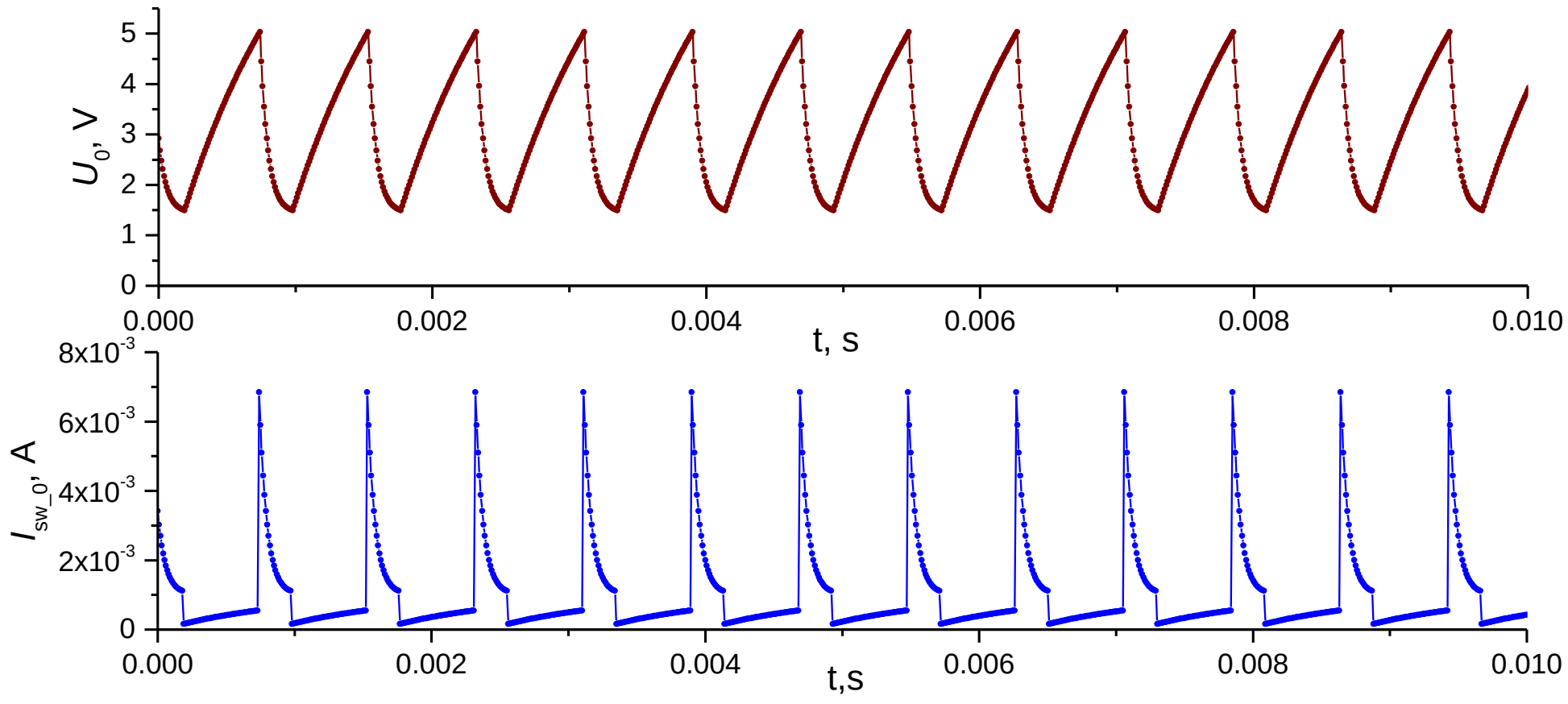

**Figure A3.** Examples of calculated oscillograms sections (1000 points), voltage $U_0$ and current $I_{sw_0}$, for the oscillator with $i = 0$ at $I_{P_0} = 1200$ μA. Here we do not show other circuit parameters because other oscillators do not affect this one.

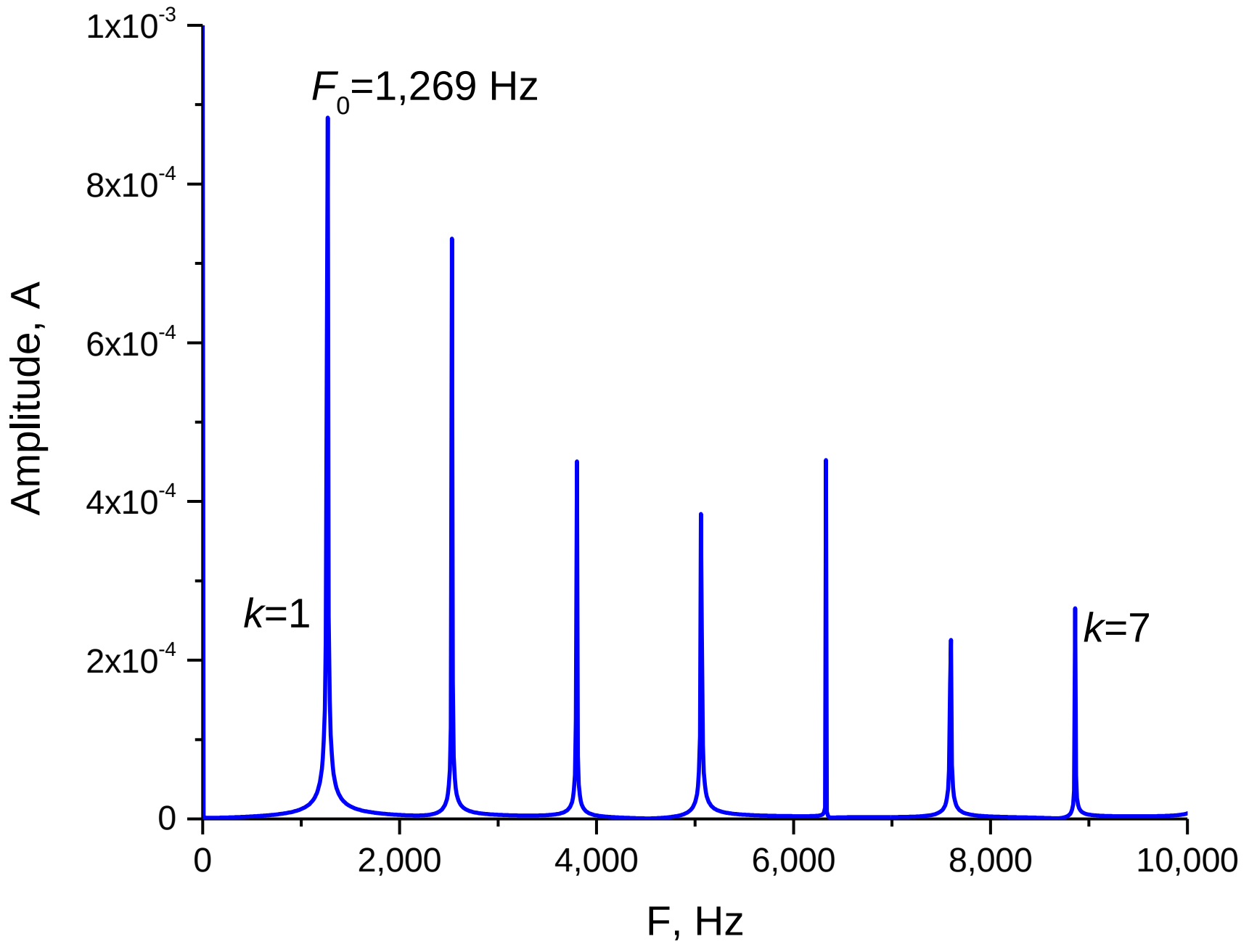

**Figure A4.** Spectrum of a current signal for the oscillogram shown in Figure A3.

Examples of sections (1000 points) of calculated oscillograms for voltage $U_0$ and current $I_{sw_0}$ are shown in Figure A3. The spectrum of the current signal has a large number of harmonics, see Figure A4. The main frequency $F^0 = 1269$ Hz determines the period of current oscillations $T^0 \sim 788$ μs (because $T^0 = 1/F^0$).

The interaction between the oscillators occurs according to the current signal as it is shown in the model, and has also been observed in an actual experiment, see [19,35], because the current signal is transformed into thermal pulses that spread along the substrate. As a result, the high order synchronization effect can be observed between the coupled oscillators.

*Appendix A.2. Dependence of $SHR_{0,10}$ and $\eta$ on the Number of Pulses in the Oscillogram*

An important issue in calculating $SHR_{0,10}$ and $\eta$ according to the method in Section 2.3, with the desired accuracy, is the determination of the number of pulses used for the calculation. In fact, it is necessary to set the minimum duration of the processed oscillogram. This is also important for determining the minimum calculation time for a family of metrics. For example, in a real experiment to calculate $SHR_{0,10}$ and $\eta$, it is required to record the oscillogram, and then make the calculation. In a model experiment, this time is determined by the oscillogram simulation time.

Figure A5a shows the dependence of $\eta$ on the number of oscillogram points with the following system parameters ($I_0$ = 1017 μA, $I_{10}$ = 891 μA, $I_{ON}$ = 725 μA, $I_{OFF}$ = 1035 μA $s_r$ = 0.1036 V, $s_m$ = 0.207 V, $s_o$ = 0.29298 V, $\eta_{th}$=90%, $U_n$ = 80 μV, the input image is $X_3$). With an increase in the number of points, the number of pulses in the oscillograms of oscillators No.0 and No.10 grows (see Figure A5b.)

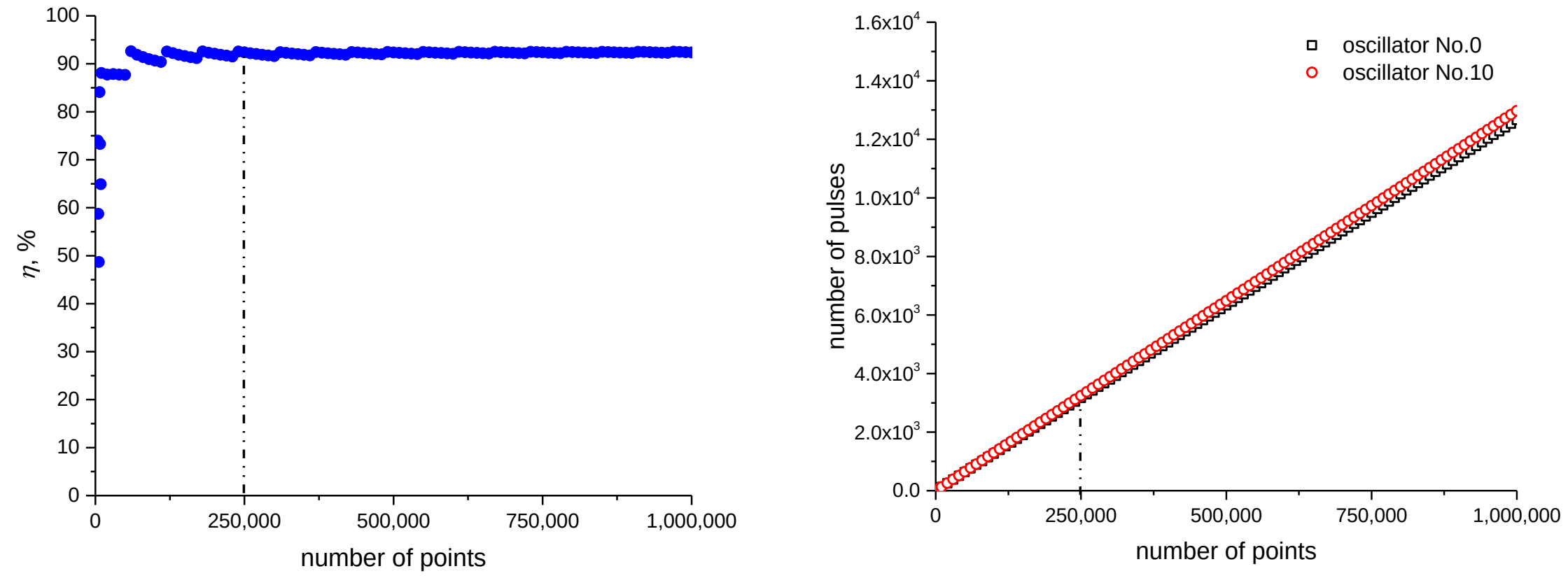

(**a**) (**b**)

**Figure A5.** The dependence of $\eta$ (**a**) and the number of pulses (**b**) on the number of points on the oscillogram. The dashed line corresponds to 250,000 oscillogram points.

With an increase in the number of points (pulses) in the processed oscillogram, the value of $\eta$ comes to saturation, which is the true value of $\eta$. The synchronization value is defined as $SHR_{0.10}$ = 38:37 with the number of points more than 60,000 (when $\eta \geq \eta_{th}$ at $\eta_{th}$ = 90%); with a smaller number of points, the synchronization is not detected ($\eta < \eta_{th}$).

In this work, we used 250,000 points (the number of pulses ~3000), which gives an error less than 0.2% in determining $\eta$. The duration of the oscillogram was ~2.5 s ($\Delta t = 10^{-5}$ s).